\documentclass[aps,prl,reprint,superscriptaddress,longbibliography]{revtex4-2}
\usepackage{graphicx} 
\usepackage{physics}
\usepackage{float}
\usepackage{lipsum}
\usepackage{graphicx}
\usepackage{amssymb}
\usepackage{natbib}
\usepackage{easyReview}
\setcitestyle{numbers}
\setcitestyle{square}
\usepackage{physics}
\setcitestyle{numbers}
\setcitestyle{square}
\usepackage{booktabs}
\usepackage{soul}

\usepackage{multirow}
\usepackage{array}
\usepackage{tabularx}
\graphicspath{{../figures/}}
\newcommand{\unity}{1\!\!1}

\usepackage[colorlinks=true,
urlcolor=blue,
linkcolor=blue,
citecolor=blue]{hyperref}
\usepackage{url}

\begin{document}
\title{Quantum resonance-enhanced performance of quantum battery}
\author{Ankita Mazumdar}
\email{ankita.m@vecc.gov.in}
\affiliation{Variable Energy Cyclotron Centre, 1/AF Bidhannagar, 
	Kolkata 
	700064, India}
\affiliation{Homi Bhabha National Institute, Training School 
	Complex, 
	Anushaktinagar, Mumbai - 400094, India}

\author{Shashi C. L. Srivastava}
\email{shashi@vecc.gov.in}
\affiliation{Variable Energy Cyclotron Centre, 1/AF Bidhannagar, 
	Kolkata 
	700064, India}
\affiliation{Homi Bhabha National Institute, Training School 
	Complex, 
	Anushaktinagar, Mumbai - 400094, India}
    
\author{Sanku Paul}
    \email{sankup005@gmail.com}
\affiliation{Department of Physics and Complex Systems, S.N. Bose 
	National 
	Centre for Basic Sciences, Kolkata 700106, India}

    \begin{abstract}
    	Quantum resonance arising whenever the ratio of the intrinsic 
    	system frequency to the driving frequency becomes a rational 
    	number has been demonstrated to generate super-linear entanglement, enhance transport, quantum metrology performance and communication. Here, we demonstrate that quantum resonance can also serve as a powerful resource for quantum batteries. We model the batteries as free 
    	rotors charged via a kicked protocol. When the individual batteries 
    	are at resonance, we show both analytically and numerically that 
    	charging power increases linearly with time while
    	efficiency 
    	(defined as the fraction of stored 
    	energy that can be extracted) remains near unity despite strong entanglement generation. 
    	Furthermore, we demonstrate that this enhanced 
    	performance persists at higher-order resonances. Demonstrating the 
    	universality of this mechanism, we show that similar enhancements 
    	arise in the interacting kicked top model, and briefly note the 
    	feasibility of its experimental realization. In a broader context, 
    	resonant charging holds significant implications for energy storage, 
    	quantum computational resources, and quantum thermodynamics.
    \end{abstract}

\maketitle

{\it Introduction$.-$} As technological scaling pushes devices increasingly toward the 
nanoscale \cite{Zhang_2007,Scully_2003,Wang_2009,Kosloff_2014,Q_sensor,Q_comp,dots, transmon, semicond, Joshi_2022}, classical paradigms of thermodynamics and energy storage are 
approaching their fundamental limits. At this microscopic frontier, energy 
transfer is no longer a continuous classical flow; instead, it is governed by 
discrete states, quantum coherence and entanglement. This transition has 
driven the conceptualization of the quantum battery \cite{Alicki_2013, Hovhannisyan_2013,Binder_2015,Campaioli_2017}, a nanoscale energy 
storage system that leverages these inherent quantum mechanical properties \cite{Kamin_2020, Arjmandi_2022,Joshi_2022,Le_2018,Ferraro_2018,Andolina_2018,Quach_2020,Zhang_2019,Andolina_2019,Zhang_2023,Shi_2022,Mula_2023,Ghosh_2020,Zhao_2021,Rossini_2019,Ghosh_2021,Francica_2017,Francica_2020,Francica_2024,Campaioli_2024,Perarnau_2015,Rosa_2020,Shaghaghi_2022,catalano_2023,Gyhm_2022,Gyhm_2024,Yang_2025,Liu_2021,Dou_2022,andolina_qm_class,gher,Gherardini,Yao_Rydberg,Grazi,romero_kickedisingQB, mazumdar_mitra_SCL2025, topological_QB} to 
store, transfer, and extract work far more efficiently than classical 
chemical counterparts. 

Consequently, recent advancements in quantum thermodynamics have heavily focused on optimizing the charging process of a quantum battery \cite{rinaldi25,Kamin_2020,shukla2025}. Most quantum battery studies focus on maintaining stable energy storage with time \cite{Rosa_2020,transmon,Santos, romero_kickedisingQB}, as coherent oscillations inevitably lead to spontaneous discharging. Consequently, the charging power decreases after reaching its maximum value, which typically occurs well before the stored energy is maximized. 
Therefore, it raises a more fundamental question : is it possible to obtain a quantum battery having a stable increase of charging power with time without any oscillation? 

Another objective in these 
optimization efforts is maximizing ergotropy 
 \cite{Allahverdyan_2004}, defined as the maximum amount 
of useful work that can be extracted from the quantum system via unitary 
processes. It has been shown that quantum 
correlations can enhance ergotropy significantly \cite{Kamin_2020, wang25_erg,yang26}.  
However, in all these systems, the efficiency—defined as the ratio of 
extracted energy to stored energy—remains strictly less than unity. In fact, maintaining 
high efficiency has proven highly problematic; internal state fluctuations 
and environmental interactions typically degrade the extractable work, 
yielding poor efficiency ratios unless the system is carefully constrained to 
operate within a near-integrable regime \cite{andolina19, Yang_2025,MitraSrivastava_2024,  mazumdar_mitra_SCL2025,Tavis_yang} or must remain in the form of product state for all time\cite{Ukhtary_23,Rodriguez_QHO,Peng_2023}.Conversely, in the chaotic regime, efficiency is generally affected negatively by the correlations present in the system \cite{Rosa_2020}.

Exploiting quantum resonance, a genuinely quantum phenomenon, we demonstrate that quantum batteries operating in a classically chaotic regime exhibits charging power that scales linearly with time while retaining near unity efficiency despite of strong entanglement generation. We model quantum batteries as free rotors. These are charged by applying time-periodic interacting potentials and tuned to quantum resonance. By deriving analytical expressions for the charging power and efficiency, in excellent agreement with numerical simulations, we reveal quantum resonance as a powerful mechanism for enhancing quantum battery performance. Owing to the unbounded Hilbert space of the interacting kicked-rotor platform, the stored energy can grow substantially without compromising efficiency. We further show that this enhancement persists even at higher-order resonances. Furthermore, we demonstrate analogous results in an interacting kicked top at resonance and propose a pathway for experimental implementation.

{\it Interacting kicked rotor platform as quantum batteries$.-$} We consider two quantum batteries, modeled by free rotors, are 
charged by 
applying a time-periodic potential consisting of kicking and interaction 
terms. In particular, the system is the interacting kicked rotor described by the Hamiltonian,
\begin{equation}
    H = \mathcal{H}_0(p_1,p_2) + V(x_1,x_2)\sum_{\tau=-\infty}^{+\infty}\delta(t-\tau T )\\
    \label{eq:IKR_ham}
\end{equation}  
where,
\begin{align*}
    \mathcal{H}_0(p_1, p_2) &= \frac{p_{1}^{2}}{2\mu_1} + \frac{p_{2}^{2}}{2\mu_2}\,,\\
    V(x_1,x_2) &= K_1\cos(x_1)+ K_2\cos(x_2)+K\cos(x_1 -x_2)\,,
\end{align*}
$x_i$, $p_i$, and $\mu_i$ denotes the position, momentum, and mass of the 
$i$th rotor, respectively. We consider unequal rotor masses, $\mu_1\neq \mu_2$, with the inverse masses satisfying $\mu_i^{-1}\in \mathbb{Z}$ ($i=1,2$). The dynamics of the individual rotors are defined 
on a cylinder with $x_i\in [0,2\pi]$ and $p_i \in (-\infty,\infty)$. In the 
potential term $V(x_1,x_2)$: $K_i$ represents the kicking strength of the 
individual rotors, while $K$ characterizes the interaction strength between 
the rotors. We set $K_i\gg 1$, so that individual rotors display 
classically chaotic dynamics. For $K=0$, the system reduces to two non-interacting kicked 
rotors which has been extensively studied \cite{IZRAILEV_1990,res1,Sanku_2016, 
Moore_1995,Sanku_2017}. 

Initially, the battery is prepared in the zero energy state, $E_{tot} 
(0)=\Tr[\mathcal{H}_0(p_1,p_2) \rho(0)]=0$,  where the density matrix 
$\rho(0) = |\psi(0)\rangle \langle \psi(0)|$ and  
$|\psi(0)\rangle=|p_1=0\rangle \otimes |p_2=0\rangle$ is the initial state. 
As the system is time-periodic, the charging is governed by the time evolution operator $U=(U_1 \otimes U_2) U_{\rm int}$, where $U_i=\exp(-ip_i^2 T/2\mu_i\hbar_s) \exp(-iK_i \cos(x_i)/\hbar_s)$ is the time-evolution operator of the $i$th rotor and $U_{\rm int}=\exp(-iK \cos(x_1-x_2)/\hbar_s)$ corresponds to the interaction term. Here, $\hbar_s$ denotes the scaled Planck's constant. The time evolution of the density matrix is then given by $\rho(t)=(U^{\dagger})^t \rho(0) U^t$. 
In general, not all of the stored energy, 
$E_{tot}(t)=\Tr[\mathcal{H}_0(p_1,p_2) \rho(t)]$ can be extracted. The amount 
of 
energy stored in the battery $\rho_i(t)$ 
is 
\begin{equation}
   E(t)= \Tr[H_0 \left(\rho_i(t) -\rho_i(0)\right)]\,,
    \label{eq:stor_ene}
\end{equation}
where $H_0=H_0(p_i)=\frac{p^2_i}{2\mu_i}$ and $\rho_i(t)$ is the 
reduced density matrix for the $i^{\text{th}}$ quantum battery.

The maximum amount of work that can be extracted from the quantum battery $\rho_i(t)$ 
through cyclic unitary transformations is given by the ergotropy \cite{Allahverdyan_2004},
\begin{equation}
    \xi(t)= \Tr{H_0 (\rho_i(t)) - \tilde{\rho}_i(t)}\,,
    \label{eq:erg}
\end{equation}
where $\tilde{\rho}_i(t)= \min_{\tilde{U}} \tilde{U}{\rho}_i(t) 
\tilde{U}^{\dagger}$ is the passive state. For $\rho_i(t) = \sum_n 
\lambda_n(t) \ket{\lambda_n(t)}\bra{\lambda_n(t)} \text{ with }
	\lambda_n(t)>\lambda_{n+1}(t)$, the passive state is given by 
	$\tilde{\rho}_i(t)= \sum_n \lambda_n(t) 
	\ket{n}\bra{n} $  and $\ket{n}$ is the eigenbasis of $H_0$ with 
	eigenvalue $\frac{n^2\hbar_s^2}{2 
		\mu_i} $.

In the non-interacting limit, $K=0$, the two rotors in presence of 
kicking term can be tuned at quantum resonance by choosing $\hbar_s T = 4\pi 
l/l'$ where $l,l'\in \mathbb{Z}$ and are co-primes.
For $l=l'=1$, system displays primary resonance, while $l,l'>1$
corresponds to higher order resonances. Note that, in the rest of the 
paper, we focus on 
	primary resonance unless higher-order 
	resonances are specified. Quantum resonance has found applications in high-precision metrology \cite{Mara_2009,Mikkel_2012}, ratchet phenomenon \cite{Lundh_2005,Dana_2008}, continuous-time quantum walk \cite{Sandro_2020}.

\begin{figure}[t]
    \includegraphics{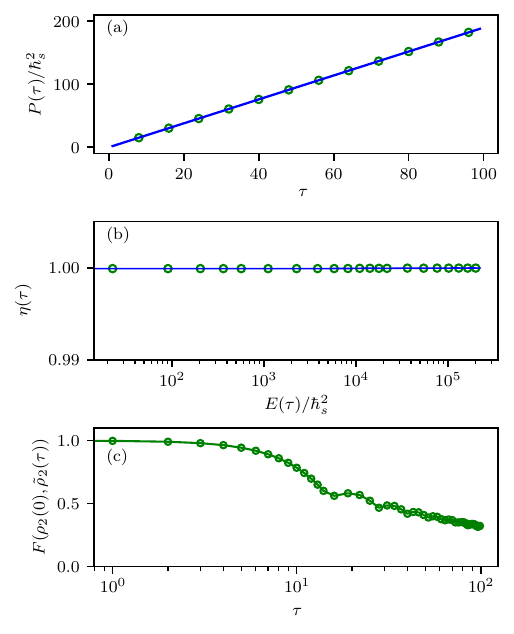}
    \caption{Plot (a) illustrates the dynamics of charging power while efficiency is shown in (b) with respect to stored energy. The plot in (c) presents the dynamics of $F(\rho_2(0),\tilde{\rho}_2(\tau))$. In (a) and (b), green circles represent numerical results whereas blue solid line corresponds to analytical results using Eq. \ref{eq_pow} and Eq. \ref{eq:efficiency}, respectively. The remaining parameters are fixed at: $K_1=9$, $K_2=10$, $K=0.1$, $\mu_1=2$, $\mu_2=1$, $L=2^{11}$ and $T=12$.}
    \label{fig:1}
\end{figure}

{\it Charging power$.-$} In the following, we analytically derive the energy storage defined in Eq. \ref{eq:stor_ene}. 
As the resonance condition is independent of $K_i$ \cite{S_paul_res}, we set $K_i=0$ without loss of generality. In addition to this, performing a coordinate transformation : $\Theta_1=x_1+x_2, ~\Theta_2=x_1-x_2,~u=\frac{(p_1+p_2)}{2},~\text{and}~v=\frac{(p_1-p_2)}{2}$, we obtain
\begin{equation}
	{E}(\tau)= \sum_{n=-L'}^{L'} \left[ J_{n} \left(\frac{K \tau}{\hbar_s}\right)\right]^2\frac{n^2 \hbar_s^2}{2}\,,
\end{equation}
where $J_n(\cdot)$ is the Bessel function of order $n$ and $\tau$ is the time in  units of $T$. 
For $K_i\neq 0$, the expression for 
${E}(\tau)$ is modified as
\begin{equation}
	{E}(\tau)= \sum_{n=-L'}^{L'}f(n,K_2,K, \tau,\hbar_s)\frac{n^2 \hbar_s^2}{2}\,,
	\label{eq:sub_ene}
\end{equation}
where $L^\prime = L/2$ with $L$ denoting the number of basis states of each rotor. Guided by numerical analysis, we find $f(n,K_2,K, \tau,\hbar_s)= J_{n} \left(\frac{K_2 \tau}{\hbar_s}\right)^2 +  J_{n} \left(\frac{K \tau}{\hbar_s}\right)^2$. 
For $L\gg 1$, $E(\tau)\approx \frac{(K_2^2 + K^2) \tau^2}{4}$ [See supplemental material]. 
Therefore, we derive  the charging power for our quantum battery to be, 
\begin{equation}
	P(\tau)\approx \frac{(K_2^2 + K^2) \tau}{4}\,.
	\label{eq_pow}
\end{equation}
Figure~\ref{fig:1}(a) shows that the analytical result is in excellent agreement with numerical simulation, further confirming that the power grows linearly with time without any added oscillations.

{\it Efficiency$.-$} We now turn to another key characteristic of a quantum battery—its efficiency. 
It can be seen from Fig.~\ref{fig:1}(b) that under resonance, nearly the entire stored energy is extractable, resulting in efficiency $\eta(\tau)=\xi(\tau)/E(\tau) \approx 1$ irrespective of the total energy stored in the quantum battery. 
This behavior highlights that quantum battery at resonance sustains near-perfect efficiency throughout the charging process.

By comparing Eqs.~\ref{eq:stor_ene} and \ref{eq:erg}, to achieve  $\eta(\tau) \approx 1$, one may assume that 
$\tilde{\rho}_i(\tau)$ remains close to 
$\rho_i(0)$, which is a pure state. However, we know from Refs. \cite{S_paul_res,Zhou_2026} at resonance, entanglement grows superlinearly with time. 
This, in fact, will prevent the passive state to have a significant overlap with the initial pure state.   This is clearly borne out by Fig.~\ref{fig:1}(c), which displays the fidelity between $\rho_2(0)$ and $\tilde{\rho}_2(\tau)$ at each time step as defined by, 
\begin{equation}
    F(\rho_2(0), \tilde{\rho}_2(\tau))= \Tr\left(\sqrt{\sqrt{\rho_2(0)}\tilde{\rho}_2(\tau)\sqrt{\rho_2(0)}} \right).
\end{equation}
In fact, at short times, the fidelity remains nearly unity, followed by  a sharp decay to a small value. This behavior where the efficiency remains close to unity despite the passive state being distinctly different from the initial state is rather counter-intuitive.

To understand this counterintuitive behavior, let us focus on the passive state energy instead of the passive state itself. The passive state energy is given by,
\begin{equation}
\tilde{E}(\tau)= \sum_n \tilde{\lambda}_n^2(\tau) \frac{n^2 \hbar_s^2}{2},
\end{equation}
where  $\tilde{\lambda}_n^2(\tau)$ are the sorted squared Scmidt eigenvalues, $\lambda_n^2(\tau)=J_n^2(K\tau/\hbar_s)$, of $\rho_i(\tau)$ for $K_i =0$ in the descending order.
 Importantly, it reveals that  
$\tilde{E}(\tau)$ is entirely governed by the interaction between the two batteries and independent of $K_i$. In contrast, as shown in Eq. \ref{eq:sub_ene}, stored energy of the quantum battery, ${E}(\tau)$, depends on both $K_2$ and $K$. The efficiency 
can be explicitly expressed as 
\begin{equation}
 \eta(\tau)=
\frac{
\sum\limits_{n=-L'}^{L'}
\left[
J_n^2\!\left(\frac{K_2\tau}{\hbar_s}\right)
+J_n^2\!\left(\frac{K\tau}{\hbar_s}\right)
-\tilde{\lambda}_n^2(\tau)
\right]
\dfrac{n^2\hbar_s^2}{2}
}{
\sum\limits_{n=-L'}^{L'}
\left[
J_n^2\!\left(\frac{K_2\tau}{\hbar_s}\right)
+J_n^2\!\left(\frac{K\tau}{\hbar_s}\right)
\right]
\dfrac{n^2\hbar_s^2}{2}
}.
\label{eq:efficiency}
\end{equation} 
In the regime $K_2\gg K$, ${E}(\tau) \gg \tilde{E}(\tau)$, which leads to 
$\eta(\tau) \approx 1$. Figure \ref{fig:1}(b) shows that the analytical 
result is in excellent agreement with the numerical simulation. This 
unambiguously proves the positive role played by resonance in 
enhancing quantum battery performance.

\begin{figure}
    \centering
    \includegraphics{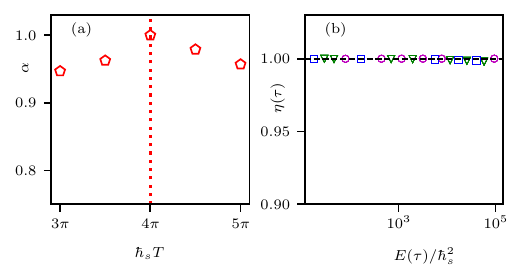}
    \caption{(a) The exponent $\alpha$ of $P(\tau)\propto \tau^{\alpha}$ is plotted with $\hbar_s T$ to observe the effect of higher order resonances. The vertical red line indicates the position of the primary resonance. (b) $\eta$ is plotted for higher order resonances by considering $\frac{l}{l^{\prime}}= \frac{3}{4}$ (magenta circles), $\frac{7}{8}$ (green lower triangles) and $\frac{9}{8}$ (blue squares). The black dashed line is guide to an eye to indicate $\eta=1$. The other parameters are same as in Fig. \ref{fig:1}. }
    \label{fig:3}
\end{figure}
{\it Higher order resonances$.-$} We address the question of whether the gain is intrinsic to the primary resonance, or if higher order resonances
can also be utilized to enhance the performance
of a quantum battery ? 
To this end, in Fig.~\ref{fig:3}(a), we plot the scaling exponent $\alpha$ of the charging 
power ($P(\tau)\propto \tau^{\alpha}$, for $\tau\gg 1$ ) as a function of $\hbar_s T$ 
to observe the effect of higher order resonances on $P(\tau)$. 
Remarkably, the stable growth of $P(\tau)$ with $\tau$ is not restricted to 
primary resonance ($\hbar_s=4\pi$). A nearly linear increase is also observed at 
higher order resonances and in particular, $\hbar_s=3\pi, 3.5\pi, 
4.5\pi,$ and $5\pi$. Similar to primary resonance, enhancement in efficiency at higher order resonances can be seen in Fig.~\ref{fig:3}(b). 
The efficiency remains nearly unity even at higher order resonances. 
This conclusively proves that quantum resonances 
of any order can be utilized to improve  quantum battery performance.

{\it Interacting kicked top and its experimental realization$.-$} Now, the 
question: is such efficient quantum battery experimentally 
amenable? The answer is affirmative. To establish a feasible platform, we 
consider a system, the kicked top, whose limiting case is the kicked rotor 
\cite{Haake_1988}. Here, we study a variant of the kicked top model, the 
interacting kicked top. We model the quantum battery as two free tops, which 
are charged by the time-dependent Hamiltonian,
\begin{equation}
\begin{aligned}
    H &= H_f + H_k \sum_{\tau=-\infty}^{+\infty}\delta(t-\tau T )\,,
\end{aligned}
\end{equation}
where
\begin{align*}
    H_k &= \sum_{m=1}^2 \alpha_m J_{m}^x + \frac{\alpha_{12}}{j} 
    J_{1}^xJ_{2}^x\,\,\, \text{and}\,\,\, 
    H_f = \sum_{m=1}^2 \frac{\beta_m}{2j} (J_{m}^{z})^2.
\end{align*}
Here $J_m=(J_{m}^x, J_{m}^y, J_{m}^z)$ ($m=1,2$) are the angular momentum operators of the $m$-th top with spin quantum number $j$. The linear term in $H_k$ proportional to $\alpha_m$ generates precession of each top about the $x$-axis, and the interaction between them is induced by the term $J_{1}^xJ_{2}^x$ with strength $\alpha_{12}$. The interaction term enables energy exchange and entanglement generation. The periodically applied quadratic term in $H_f$ proportional to $\beta_m$ produces the nonlinear torsion responsible for chaotic dynamics of each top.  Throughout this work, we set $\hbar=1$.

Now, to investigate the effect of resonance on the battery performance composed of two tops, we impose the primary resonance condition by setting $\beta_m=4\pi j$. The system is initialized in the fully polarized state $|\psi(0) = |J_{1}^z=-j\rangle \otimes |J_{2}^z=-j\rangle$. The charging is performed by applying the period-$1$ time evolution operator $U=\exp( -iH_f) \exp( -iH_k)$. The time-evolved state is obtained as $|\psi(t)\rangle = U^t |\psi(0) \rangle$. Unlike the kicked rotor, the resulting system evolves in a finite-dimensional Hilbert space that provides a restriction on the maximum energy storage.

\begin{figure}[t]
    \centering
    \includegraphics{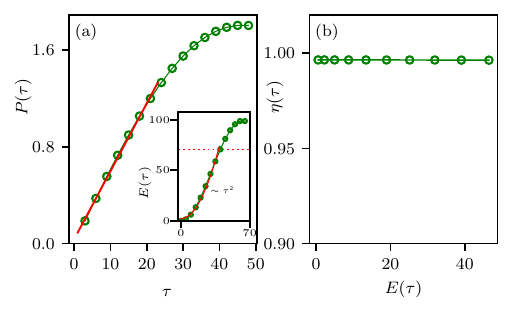}
    \caption{(a) The dynamics of $P(\tau)$ is shown, with the red solid line indicating a linear fit. The inset displays $E(\tau)$, where the numerical data are fitted by $E(\tau) \sim \tau^2$ (red solid line). The horizontal red-dashed line marks $E(\tau) =70.66$ i.e. almost $70\%$ of the bandwidth, upto which the quadratic growth persists. 
    (b) The $\eta(\tau)$ is plotted with respect to $E(\tau)$. Regardless of the growth of energy, $\eta(\tau) \approx 1$. The parameters are set as $j=50, \alpha_1= 1,\alpha_2= 0.05, \alpha_{12}= 0.03$.}
    \label{fig:top}
\end{figure}

Figure~\ref{fig:top} shows charging power and efficiency of the quantum battery, with quantum battery Hamiltonian, $H_B=J_1^z$. It can be seen in Fig. \ref{fig:top}(a) that $P(\tau)$ scales linearly with time until $E(\tau)/E_{\rm max}<0.71$, where $E_{\rm max}=2j$. Remarkably, it maintains $\eta(\tau)\approx 1$ throughout the charging process (Fig. \ref{fig:top}(b)) even though the system displays strong entanglement generation [See supplemental material]. These results reveal that the resonance-enhanced quantum battery performance survives the compactification of the rotor dynamics. 

Experimentally, the kicked top has been realized in a variety of 
collective-spin platforms, including ultracold atomic ensembles 
\cite{Jessen_2009}, superconducting qubits \cite{Neill_2016}, nuclear 
magnetic resonance (NMR) \cite{Krithika_2019}. In particular, the interaction 
$J_1^xJ_2^x$ can be engineered through cavity-mediated coherent spin-exchange 
processes \cite{Thompson_2018}, providing a route for experimental 
implementation of resonance-enhanced quantum batteries. It is worth 
mentioning that experiments on quantum batteries based on harmonic 
oscillators have demonstrated the storage of an arbitrary amount of energy 
\cite{catalyst_QHO,Peng_2023}, highlighting the relevance of such infinite 
dimensional models.

{\it Conclusion and outlook$.-$} In summary, we have demonstrated the positive role of quantum 
resonance in enhancing quantum battery performance, even when the system 
operates in a classically chaotic regime and under strong entanglement production. Modeled by free rotors and charged using 
kicked protocols, the quantum battery exhibits quadratic energy growth without any oscillation and consequently a
linear increase of charging power with time when individual batteries are in 
resonance during 
charging. Notably, this maintains near-perfect efficiency, meaning all stored 
energy can be extracted even in the presence of strong entanglement 
production. Under resonant condition we analytically establish that this 
phenomenon originates from the fact that the passive energy becomes 
independent of the individual kick strength and only depends upon the 
interaction strength. In contrast, the stored energy depends upon the kick 
strength of the subsystem and the interaction strength. Thus, resonance 
serves as a powerful tool to mitigate the negative role played by the 
entanglement. We have also demonstrated the 
persistence of this enhanced performance at higher-order resonances. To 
establish the broader universality of this mechanism, we extended our 
analysis to an interacting kicked top model, demonstrating that it exhibits 
identical resonant enhancements. Furthermore, we provide a viable 
experimental outlook for such systems, noting that the requisite individual 
operations have already been successfully realized in distinct experimental 
settings. These 
findings open up several new research directions, including exploring the 
role of quantum resonance in coupled batteries and investigating potential 
quantum advantages \cite{andolina_25}.

\begin{acknowledgments}
S.P. would like to thank DST India for the Inspire Faculty
Grant.
\end{acknowledgments}

\bibliographystyle{apsrev4-2}
\bibliography{ref_rotor}

\begin{thebibliography}{81}%
\makeatletter
\providecommand \@ifxundefined [1]{%
 \@ifx{#1\undefined}
}%
\providecommand \@ifnum [1]{%
 \ifnum #1\expandafter \@firstoftwo
 \else \expandafter \@secondoftwo
 \fi
}%
\providecommand \@ifx [1]{%
 \ifx #1\expandafter \@firstoftwo
 \else \expandafter \@secondoftwo
 \fi
}%
\providecommand \natexlab [1]{#1}%
\providecommand \enquote  [1]{``#1''}%
\providecommand \bibnamefont  [1]{#1}%
\providecommand \bibfnamefont [1]{#1}%
\providecommand \citenamefont [1]{#1}%
\providecommand \href@noop [0]{\@secondoftwo}%
\providecommand \href [0]{\begingroup \@sanitize@url \@href}%
\providecommand \@href[1]{\@@startlink{#1}\@@href}%
\providecommand \@@href[1]{\endgroup#1\@@endlink}%
\providecommand \@sanitize@url [0]{\catcode `\\12\catcode `\$12\catcode
  `\&12\catcode `\#12\catcode `\^12\catcode `\_12\catcode `\%12\relax}%
\providecommand \@@startlink[1]{}%
\providecommand \@@endlink[0]{}%
\providecommand \url  [0]{\begingroup\@sanitize@url \@url }%
\providecommand \@url [1]{\endgroup\@href {#1}{\urlprefix }}%
\providecommand \urlprefix  [0]{URL }%
\providecommand \Eprint [0]{\href }%
\providecommand \doibase [0]{https://doi.org/}%
\providecommand \selectlanguage [0]{\@gobble}%
\providecommand \bibinfo  [0]{\@secondoftwo}%
\providecommand \bibfield  [0]{\@secondoftwo}%
\providecommand \translation [1]{[#1]}%
\providecommand \BibitemOpen [0]{}%
\providecommand \bibitemStop [0]{}%
\providecommand \bibitemNoStop [0]{.\EOS\space}%
\providecommand \EOS [0]{\spacefactor3000\relax}%
\providecommand \BibitemShut  [1]{\csname bibitem#1\endcsname}%
\let\auto@bib@innerbib\@empty
\bibitem [{\citenamefont {Zhang}\ \emph {et~al.}(2007)\citenamefont {Zhang},
  \citenamefont {Liu}, \citenamefont {Chen},\ and\ \citenamefont
  {Li}}]{Zhang_2007}%
  \BibitemOpen
  \bibfield  {author} {\bibinfo {author} {\bibfnamefont {T.}~\bibnamefont
  {Zhang}}, \bibinfo {author} {\bibfnamefont {W.-T.}\ \bibnamefont {Liu}},
  \bibinfo {author} {\bibfnamefont {P.-X.}\ \bibnamefont {Chen}},\ and\
  \bibinfo {author} {\bibfnamefont {C.-Z.}\ \bibnamefont {Li}},\ }\href
  {https://doi.org/10.1103/PhysRevA.75.062102} {\bibfield  {journal} {\bibinfo
  {journal} {Phys. Rev. A}\ }\textbf {\bibinfo {volume} {75}},\ \bibinfo
  {pages} {062102} (\bibinfo {year} {2007})}\BibitemShut {NoStop}%
\bibitem [{\citenamefont {Scully}\ \emph {et~al.}(2003)\citenamefont {Scully},
  \citenamefont {Zubairy}, \citenamefont {Agarwal},\ and\ \citenamefont
  {Walther}}]{Scully_2003}%
  \BibitemOpen
  \bibfield  {author} {\bibinfo {author} {\bibfnamefont {M.~O.}\ \bibnamefont
  {Scully}}, \bibinfo {author} {\bibfnamefont {M.~S.}\ \bibnamefont {Zubairy}},
  \bibinfo {author} {\bibfnamefont {G.~S.}\ \bibnamefont {Agarwal}},\ and\
  \bibinfo {author} {\bibfnamefont {H.}~\bibnamefont {Walther}},\ }\href
  {https://doi.org/10.1126/science.1078955} {\bibfield  {journal} {\bibinfo
  {journal} {Science}\ }\textbf {\bibinfo {volume} {299}},\ \bibinfo {pages}
  {862} (\bibinfo {year} {2003})}\BibitemShut {NoStop}%
\bibitem [{\citenamefont {Wang}\ \emph {et~al.}(2009)\citenamefont {Wang},
  \citenamefont {Liu},\ and\ \citenamefont {He}}]{Wang_2009}%
  \BibitemOpen
  \bibfield  {author} {\bibinfo {author} {\bibfnamefont {H.}~\bibnamefont
  {Wang}}, \bibinfo {author} {\bibfnamefont {S.}~\bibnamefont {Liu}},\ and\
  \bibinfo {author} {\bibfnamefont {J.}~\bibnamefont {He}},\ }\href
  {https://doi.org/10.1103/PhysRevE.79.041113} {\bibfield  {journal} {\bibinfo
  {journal} {Phys. Rev. E}\ }\textbf {\bibinfo {volume} {79}},\ \bibinfo
  {pages} {041113} (\bibinfo {year} {2009})}\BibitemShut {NoStop}%
\bibitem [{\citenamefont {Kosloff}\ and\ \citenamefont
  {Levy}(2014)}]{Kosloff_2014}%
  \BibitemOpen
  \bibfield  {author} {\bibinfo {author} {\bibfnamefont {R.}~\bibnamefont
  {Kosloff}}\ and\ \bibinfo {author} {\bibfnamefont {A.}~\bibnamefont {Levy}},\
  }\href
  {https://doi.org/https://doi.org/10.1146/annurev-physchem-040513-103724}
  {\bibfield  {journal} {\bibinfo  {journal} {Annu. Rev. Phys. Chem.}\ }\textbf
  {\bibinfo {volume} {65}},\ \bibinfo {pages} {365} (\bibinfo {year}
  {2014})}\BibitemShut {NoStop}%
\bibitem [{\citenamefont {Degen}\ \emph {et~al.}(2017)\citenamefont {Degen},
  \citenamefont {Reinhard},\ and\ \citenamefont {Cappellaro}}]{Q_sensor}%
  \BibitemOpen
  \bibfield  {author} {\bibinfo {author} {\bibfnamefont {C.~L.}\ \bibnamefont
  {Degen}}, \bibinfo {author} {\bibfnamefont {F.}~\bibnamefont {Reinhard}},\
  and\ \bibinfo {author} {\bibfnamefont {P.}~\bibnamefont {Cappellaro}},\
  }\href {https://doi.org/10.1103/RevModPhys.89.035002} {\bibfield  {journal}
  {\bibinfo  {journal} {Rev. Mod. Phys.}\ }\textbf {\bibinfo {volume} {89}},\
  \bibinfo {pages} {035002} (\bibinfo {year} {2017})}\BibitemShut {NoStop}%
\bibitem [{\citenamefont {P\'erez-Delgado}\ and\ \citenamefont
  {Kok}(2011)}]{Q_comp}%
  \BibitemOpen
  \bibfield  {author} {\bibinfo {author} {\bibfnamefont {C.~A.}\ \bibnamefont
  {P\'erez-Delgado}}\ and\ \bibinfo {author} {\bibfnamefont {P.}~\bibnamefont
  {Kok}},\ }\href {https://doi.org/10.1103/PhysRevA.83.012303} {\bibfield
  {journal} {\bibinfo  {journal} {Phys. Rev. A}\ }\textbf {\bibinfo {volume}
  {83}},\ \bibinfo {pages} {012303} (\bibinfo {year} {2011})}\BibitemShut
  {NoStop}%
\bibitem [{\citenamefont {Maillette~de Buy~Wenniger}\ \emph
  {et~al.}(2023)\citenamefont {Maillette~de Buy~Wenniger}, \citenamefont
  {Thomas}, \citenamefont {Maffei}, \citenamefont {Wein}, \citenamefont {Pont},
  \citenamefont {Belabas}, \citenamefont {Prasad}, \citenamefont {Harouri},
  \citenamefont {Lema\^{\i}tre}, \citenamefont {Sagnes}, \citenamefont
  {Somaschi}, \citenamefont {Auff\`eves},\ and\ \citenamefont
  {Senellart}}]{dots}%
  \BibitemOpen
  \bibfield  {author} {\bibinfo {author} {\bibfnamefont {I.}~\bibnamefont
  {Maillette~de Buy~Wenniger}}, \bibinfo {author} {\bibfnamefont {S.~E.}\
  \bibnamefont {Thomas}}, \bibinfo {author} {\bibfnamefont {M.}~\bibnamefont
  {Maffei}}, \bibinfo {author} {\bibfnamefont {S.~C.}\ \bibnamefont {Wein}},
  \bibinfo {author} {\bibfnamefont {M.}~\bibnamefont {Pont}}, \bibinfo {author}
  {\bibfnamefont {N.}~\bibnamefont {Belabas}}, \bibinfo {author} {\bibfnamefont
  {S.}~\bibnamefont {Prasad}}, \bibinfo {author} {\bibfnamefont
  {A.}~\bibnamefont {Harouri}}, \bibinfo {author} {\bibfnamefont
  {A.}~\bibnamefont {Lema\^{\i}tre}}, \bibinfo {author} {\bibfnamefont
  {I.}~\bibnamefont {Sagnes}}, \bibinfo {author} {\bibfnamefont
  {N.}~\bibnamefont {Somaschi}}, \bibinfo {author} {\bibfnamefont
  {A.}~\bibnamefont {Auff\`eves}},\ and\ \bibinfo {author} {\bibfnamefont
  {P.}~\bibnamefont {Senellart}},\ }\href
  {https://doi.org/10.1103/PhysRevLett.131.260401} {\bibfield  {journal}
  {\bibinfo  {journal} {Phys. Rev. Lett.}\ }\textbf {\bibinfo {volume} {131}},\
  \bibinfo {pages} {260401} (\bibinfo {year} {2023})}\BibitemShut {NoStop}%
\bibitem [{\citenamefont {Dou}\ and\ \citenamefont {Yang}(2023)}]{transmon}%
  \BibitemOpen
  \bibfield  {author} {\bibinfo {author} {\bibfnamefont {F.-Q.}\ \bibnamefont
  {Dou}}\ and\ \bibinfo {author} {\bibfnamefont {F.-M.}\ \bibnamefont {Yang}},\
  }\href {https://doi.org/10.1103/PhysRevA.107.023725} {\bibfield  {journal}
  {\bibinfo  {journal} {Phys. Rev. A}\ }\textbf {\bibinfo {volume} {107}},\
  \bibinfo {pages} {023725} (\bibinfo {year} {2023})}\BibitemShut {NoStop}%
\bibitem [{\citenamefont {Quach}\ \emph {et~al.}(2022)\citenamefont {Quach},
  \citenamefont {McGhee}, \citenamefont {Ganzer}, \citenamefont {Rouse},
  \citenamefont {Lovett}, \citenamefont {Gauger}, \citenamefont {Keeling},
  \citenamefont {Cerullo}, \citenamefont {Lidzey},\ and\ \citenamefont
  {Virgili}}]{semicond}%
  \BibitemOpen
  \bibfield  {author} {\bibinfo {author} {\bibfnamefont {J.~Q.}\ \bibnamefont
  {Quach}}, \bibinfo {author} {\bibfnamefont {K.~E.}\ \bibnamefont {McGhee}},
  \bibinfo {author} {\bibfnamefont {L.}~\bibnamefont {Ganzer}}, \bibinfo
  {author} {\bibfnamefont {D.~M.}\ \bibnamefont {Rouse}}, \bibinfo {author}
  {\bibfnamefont {B.~W.}\ \bibnamefont {Lovett}}, \bibinfo {author}
  {\bibfnamefont {E.~M.}\ \bibnamefont {Gauger}}, \bibinfo {author}
  {\bibfnamefont {J.}~\bibnamefont {Keeling}}, \bibinfo {author} {\bibfnamefont
  {G.}~\bibnamefont {Cerullo}}, \bibinfo {author} {\bibfnamefont {D.~G.}\
  \bibnamefont {Lidzey}},\ and\ \bibinfo {author} {\bibfnamefont
  {T.}~\bibnamefont {Virgili}},\ }\href
  {https://doi.org/10.1126/sciadv.abk3160} {\bibfield  {journal} {\bibinfo
  {journal} {Sci. Adv.}\ }\textbf {\bibinfo {volume} {8}},\ \bibinfo {pages}
  {eabk3160} (\bibinfo {year} {2022})}\BibitemShut {NoStop}%
\bibitem [{\citenamefont {Joshi}\ and\ \citenamefont
  {Mahesh}(2022)}]{Joshi_2022}%
  \BibitemOpen
  \bibfield  {author} {\bibinfo {author} {\bibfnamefont {J.}~\bibnamefont
  {Joshi}}\ and\ \bibinfo {author} {\bibfnamefont {T.~S.}\ \bibnamefont
  {Mahesh}},\ }\href {https://doi.org/10.1103/PhysRevA.106.042601} {\bibfield
  {journal} {\bibinfo  {journal} {Phys. Rev. A}\ }\textbf {\bibinfo {volume}
  {106}},\ \bibinfo {pages} {042601} (\bibinfo {year} {2022})}\BibitemShut
  {NoStop}%
\bibitem [{\citenamefont {Alicki}\ and\ \citenamefont
  {Fannes}(2013)}]{Alicki_2013}%
  \BibitemOpen
  \bibfield  {author} {\bibinfo {author} {\bibfnamefont {R.}~\bibnamefont
  {Alicki}}\ and\ \bibinfo {author} {\bibfnamefont {M.}~\bibnamefont
  {Fannes}},\ }\href {https://doi.org/10.1103/PhysRevE.87.042123} {\bibfield
  {journal} {\bibinfo  {journal} {Phys. Rev. E}\ }\textbf {\bibinfo {volume}
  {87}},\ \bibinfo {pages} {042123} (\bibinfo {year} {2013})}\BibitemShut
  {NoStop}%
\bibitem [{\citenamefont {Hovhannisyan}\ \emph {et~al.}(2013)\citenamefont
  {Hovhannisyan}, \citenamefont {Perarnau-Llobet}, \citenamefont {Huber},\ and\
  \citenamefont {Ac\'{\i}n}}]{Hovhannisyan_2013}%
  \BibitemOpen
  \bibfield  {author} {\bibinfo {author} {\bibfnamefont {K.~V.}\ \bibnamefont
  {Hovhannisyan}}, \bibinfo {author} {\bibfnamefont {M.}~\bibnamefont
  {Perarnau-Llobet}}, \bibinfo {author} {\bibfnamefont {M.}~\bibnamefont
  {Huber}},\ and\ \bibinfo {author} {\bibfnamefont {A.}~\bibnamefont
  {Ac\'{\i}n}},\ }\href {https://doi.org/10.1103/PhysRevLett.111.240401}
  {\bibfield  {journal} {\bibinfo  {journal} {Phys. Rev. Lett.}\ }\textbf
  {\bibinfo {volume} {111}},\ \bibinfo {pages} {240401} (\bibinfo {year}
  {2013})}\BibitemShut {NoStop}%
\bibitem [{\citenamefont {Binder}\ \emph {et~al.}(2015)\citenamefont {Binder},
  \citenamefont {Vinjanampathy}, \citenamefont {Modi},\ and\ \citenamefont
  {Goold}}]{Binder_2015}%
  \BibitemOpen
  \bibfield  {author} {\bibinfo {author} {\bibfnamefont {F.~C.}\ \bibnamefont
  {Binder}}, \bibinfo {author} {\bibfnamefont {S.}~\bibnamefont
  {Vinjanampathy}}, \bibinfo {author} {\bibfnamefont {K.}~\bibnamefont
  {Modi}},\ and\ \bibinfo {author} {\bibfnamefont {J.}~\bibnamefont {Goold}},\
  }\href {https://doi.org/10.1088/1367-2630/17/7/075015} {\bibfield  {journal}
  {\bibinfo  {journal} {New J. Phys.}\ }\textbf {\bibinfo {volume} {17}},\
  \bibinfo {pages} {075015} (\bibinfo {year} {2015})}\BibitemShut {NoStop}%
\bibitem [{\citenamefont {Campaioli}\ \emph {et~al.}()\citenamefont
  {Campaioli}, \citenamefont {Pollock}, \citenamefont {Binder}, \citenamefont
  {C{\'e}leri}, \citenamefont {Goold}, \citenamefont {Vinjanampathy},\ and\
  \citenamefont {Modi}}]{Campaioli_2017}%
  \BibitemOpen
  \bibfield  {author} {\bibinfo {author} {\bibfnamefont {F.}~\bibnamefont
  {Campaioli}}, \bibinfo {author} {\bibfnamefont {F.}~\bibnamefont {Pollock}},
  \bibinfo {author} {\bibfnamefont {F.}~\bibnamefont {Binder}}, \bibinfo
  {author} {\bibfnamefont {L.}~\bibnamefont {C{\'e}leri}}, \bibinfo {author}
  {\bibfnamefont {J.}~\bibnamefont {Goold}}, \bibinfo {author} {\bibfnamefont
  {S.}~\bibnamefont {Vinjanampathy}},\ and\ \bibinfo {author} {\bibfnamefont
  {K.}~\bibnamefont {Modi}},\ }\href
  {https://doi.org/10.1103/PhysRevLett.118.150601} {\bibfield  {journal}
  {\bibinfo  {journal} {Phys. Rev. Lett.}\ }\textbf {\bibinfo {volume} {118}},\
  \bibinfo {pages} {150601}}\BibitemShut {NoStop}%
\bibitem [{\citenamefont {Kamin}\ \emph {et~al.}(2020)\citenamefont {Kamin},
  \citenamefont {Tabesh}, \citenamefont {Salimi},\ and\ \citenamefont
  {Santos}}]{Kamin_2020}%
  \BibitemOpen
  \bibfield  {author} {\bibinfo {author} {\bibfnamefont {F.~H.}\ \bibnamefont
  {Kamin}}, \bibinfo {author} {\bibfnamefont {F.~T.}\ \bibnamefont {Tabesh}},
  \bibinfo {author} {\bibfnamefont {S.}~\bibnamefont {Salimi}},\ and\ \bibinfo
  {author} {\bibfnamefont {A.~C.}\ \bibnamefont {Santos}},\ }\href
  {https://doi.org/10.1103/PhysRevE.102.052109} {\bibfield  {journal} {\bibinfo
   {journal} {Phys. Rev. E}\ }\textbf {\bibinfo {volume} {102}},\ \bibinfo
  {pages} {052109} (\bibinfo {year} {2020})}\BibitemShut {NoStop}%
\bibitem [{\citenamefont {Arjmandi}\ \emph {et~al.}(2022)\citenamefont
  {Arjmandi}, \citenamefont {Shokri}, \citenamefont {Faizi},\ and\
  \citenamefont {Mohammadi}}]{Arjmandi_2022}%
  \BibitemOpen
  \bibfield  {author} {\bibinfo {author} {\bibfnamefont {M.~B.}\ \bibnamefont
  {Arjmandi}}, \bibinfo {author} {\bibfnamefont {A.}~\bibnamefont {Shokri}},
  \bibinfo {author} {\bibfnamefont {E.}~\bibnamefont {Faizi}},\ and\ \bibinfo
  {author} {\bibfnamefont {H.}~\bibnamefont {Mohammadi}},\ }\href
  {https://doi.org/10.1103/PhysRevA.106.062609} {\bibfield  {journal} {\bibinfo
   {journal} {Phys. Rev. A}\ }\textbf {\bibinfo {volume} {106}},\ \bibinfo
  {pages} {062609} (\bibinfo {year} {2022})}\BibitemShut {NoStop}%
\bibitem [{\citenamefont {Le}\ \emph {et~al.}(2018)\citenamefont {Le},
  \citenamefont {Levinsen}, \citenamefont {Modi}, \citenamefont {Parish},\ and\
  \citenamefont {Pollock}}]{Le_2018}%
  \BibitemOpen
  \bibfield  {author} {\bibinfo {author} {\bibfnamefont {T.~P.}\ \bibnamefont
  {Le}}, \bibinfo {author} {\bibfnamefont {J.}~\bibnamefont {Levinsen}},
  \bibinfo {author} {\bibfnamefont {K.}~\bibnamefont {Modi}}, \bibinfo {author}
  {\bibfnamefont {M.~M.}\ \bibnamefont {Parish}},\ and\ \bibinfo {author}
  {\bibfnamefont {F.~A.}\ \bibnamefont {Pollock}},\ }\href
  {https://doi.org/10.1103/PhysRevA.97.022106} {\bibfield  {journal} {\bibinfo
  {journal} {Phys. Rev. A}\ }\textbf {\bibinfo {volume} {97}},\ \bibinfo
  {pages} {022106} (\bibinfo {year} {2018})}\BibitemShut {NoStop}%
\bibitem [{\citenamefont {Ferraro}\ \emph {et~al.}(2018)\citenamefont
  {Ferraro}, \citenamefont {Campisi}, \citenamefont {Andolina}, \citenamefont
  {Pellegrini},\ and\ \citenamefont {Polini}}]{Ferraro_2018}%
  \BibitemOpen
  \bibfield  {author} {\bibinfo {author} {\bibfnamefont {D.}~\bibnamefont
  {Ferraro}}, \bibinfo {author} {\bibfnamefont {M.}~\bibnamefont {Campisi}},
  \bibinfo {author} {\bibfnamefont {G.~M.}\ \bibnamefont {Andolina}}, \bibinfo
  {author} {\bibfnamefont {V.}~\bibnamefont {Pellegrini}},\ and\ \bibinfo
  {author} {\bibfnamefont {M.}~\bibnamefont {Polini}},\ }\href
  {https://doi.org/10.1103/PhysRevLett.120.117702} {\bibfield  {journal}
  {\bibinfo  {journal} {Phys. Rev. Lett.}\ }\textbf {\bibinfo {volume} {120}},\
  \bibinfo {pages} {117702} (\bibinfo {year} {2018})}\BibitemShut {NoStop}%
\bibitem [{\citenamefont {Andolina}\ \emph {et~al.}(2018)\citenamefont
  {Andolina}, \citenamefont {Farina}, \citenamefont {Mari}, \citenamefont
  {Pellegrini}, \citenamefont {Giovannetti},\ and\ \citenamefont
  {Polini}}]{Andolina_2018}%
  \BibitemOpen
  \bibfield  {author} {\bibinfo {author} {\bibfnamefont {G.~M.}\ \bibnamefont
  {Andolina}}, \bibinfo {author} {\bibfnamefont {D.}~\bibnamefont {Farina}},
  \bibinfo {author} {\bibfnamefont {A.}~\bibnamefont {Mari}}, \bibinfo {author}
  {\bibfnamefont {V.}~\bibnamefont {Pellegrini}}, \bibinfo {author}
  {\bibfnamefont {V.}~\bibnamefont {Giovannetti}},\ and\ \bibinfo {author}
  {\bibfnamefont {M.}~\bibnamefont {Polini}},\ }\href
  {https://doi.org/10.1103/PhysRevB.98.205423} {\bibfield  {journal} {\bibinfo
  {journal} {Phys. Rev. B}\ }\textbf {\bibinfo {volume} {98}},\ \bibinfo
  {pages} {205423} (\bibinfo {year} {2018})}\BibitemShut {NoStop}%
\bibitem [{\citenamefont {Quach}\ and\ \citenamefont
  {Munro}(2020)}]{Quach_2020}%
  \BibitemOpen
  \bibfield  {author} {\bibinfo {author} {\bibfnamefont {J.~Q.}\ \bibnamefont
  {Quach}}\ and\ \bibinfo {author} {\bibfnamefont {W.~J.}\ \bibnamefont
  {Munro}},\ }\href {https://doi.org/10.1103/PhysRevApplied.14.024092}
  {\bibfield  {journal} {\bibinfo  {journal} {Phys. Rev. Appl.}\ }\textbf
  {\bibinfo {volume} {14}},\ \bibinfo {pages} {024092} (\bibinfo {year}
  {2020})}\BibitemShut {NoStop}%
\bibitem [{\citenamefont {Zhang}\ \emph {et~al.}(2019)\citenamefont {Zhang},
  \citenamefont {Yang}, \citenamefont {Fu},\ and\ \citenamefont
  {Wang}}]{Zhang_2019}%
  \BibitemOpen
  \bibfield  {author} {\bibinfo {author} {\bibfnamefont {Y.-Y.}\ \bibnamefont
  {Zhang}}, \bibinfo {author} {\bibfnamefont {T.-R.}\ \bibnamefont {Yang}},
  \bibinfo {author} {\bibfnamefont {L.}~\bibnamefont {Fu}},\ and\ \bibinfo
  {author} {\bibfnamefont {X.}~\bibnamefont {Wang}},\ }\href
  {https://doi.org/10.1103/PhysRevE.99.052106} {\bibfield  {journal} {\bibinfo
  {journal} {Phys. Rev. E}\ }\textbf {\bibinfo {volume} {99}},\ \bibinfo
  {pages} {052106} (\bibinfo {year} {2019})}\BibitemShut {NoStop}%
\bibitem [{\citenamefont {Andolina}\ \emph
  {et~al.}(2019{\natexlab{a}})\citenamefont {Andolina}, \citenamefont {Keck},
  \citenamefont {Mari}, \citenamefont {Campisi}, \citenamefont {Giovannetti},\
  and\ \citenamefont {Polini}}]{Andolina_2019}%
  \BibitemOpen
  \bibfield  {author} {\bibinfo {author} {\bibfnamefont {G.~M.}\ \bibnamefont
  {Andolina}}, \bibinfo {author} {\bibfnamefont {M.}~\bibnamefont {Keck}},
  \bibinfo {author} {\bibfnamefont {A.}~\bibnamefont {Mari}}, \bibinfo {author}
  {\bibfnamefont {M.}~\bibnamefont {Campisi}}, \bibinfo {author} {\bibfnamefont
  {V.}~\bibnamefont {Giovannetti}},\ and\ \bibinfo {author} {\bibfnamefont
  {M.}~\bibnamefont {Polini}},\ }\href
  {https://doi.org/10.1103/PhysRevLett.122.047702} {\bibfield  {journal}
  {\bibinfo  {journal} {Phys. Rev. Lett.}\ }\textbf {\bibinfo {volume} {122}},\
  \bibinfo {pages} {047702} (\bibinfo {year} {2019}{\natexlab{a}})}\BibitemShut
  {NoStop}%
\bibitem [{\citenamefont {Zhang}\ and\ \citenamefont
  {Blaauboer}(2023)}]{Zhang_2023}%
  \BibitemOpen
  \bibfield  {author} {\bibinfo {author} {\bibfnamefont {X.}~\bibnamefont
  {Zhang}}\ and\ \bibinfo {author} {\bibfnamefont {M.}~\bibnamefont
  {Blaauboer}},\ }\href {https://doi.org/10.3389/fphy.2022.1097564} {\bibfield
  {journal} {\bibinfo  {journal} {Front. Phys.}\ }\textbf {\bibinfo {volume}
  {10}},\ \bibinfo {pages} {1097564} (\bibinfo {year} {2023})}\BibitemShut
  {NoStop}%
\bibitem [{\citenamefont {Shi}\ \emph {et~al.}(2022)\citenamefont {Shi},
  \citenamefont {Ding}, \citenamefont {Wan}, \citenamefont {Wang},\ and\
  \citenamefont {Yang}}]{Shi_2022}%
  \BibitemOpen
  \bibfield  {author} {\bibinfo {author} {\bibfnamefont {H.-L.}\ \bibnamefont
  {Shi}}, \bibinfo {author} {\bibfnamefont {S.}~\bibnamefont {Ding}}, \bibinfo
  {author} {\bibfnamefont {Q.-K.}\ \bibnamefont {Wan}}, \bibinfo {author}
  {\bibfnamefont {X.-H.}\ \bibnamefont {Wang}},\ and\ \bibinfo {author}
  {\bibfnamefont {W.-L.}\ \bibnamefont {Yang}},\ }\href
  {https://doi.org/10.1103/PhysRevLett.129.130602} {\bibfield  {journal}
  {\bibinfo  {journal} {Phys. Rev. Lett.}\ }\textbf {\bibinfo {volume} {129}},\
  \bibinfo {pages} {130602} (\bibinfo {year} {2022})}\BibitemShut {NoStop}%
\bibitem [{\citenamefont {Mula}\ \emph {et~al.}(2023)\citenamefont {Mula},
  \citenamefont {Fern\'andez}, \citenamefont {Alvarellos}, \citenamefont
  {Fern\'andez}, \citenamefont {Garc\'{\i}a-Aldea}, \citenamefont {Santalla},\
  and\ \citenamefont {Rodr\'{\i}guez-Laguna}}]{Mula_2023}%
  \BibitemOpen
  \bibfield  {author} {\bibinfo {author} {\bibfnamefont {B.~n.}\ \bibnamefont
  {Mula}}, \bibinfo {author} {\bibfnamefont {E.~M.}\ \bibnamefont
  {Fern\'andez}}, \bibinfo {author} {\bibfnamefont {J.~E.}\ \bibnamefont
  {Alvarellos}}, \bibinfo {author} {\bibfnamefont {J.~J.}\ \bibnamefont
  {Fern\'andez}}, \bibinfo {author} {\bibfnamefont {D.}~\bibnamefont
  {Garc\'{\i}a-Aldea}}, \bibinfo {author} {\bibfnamefont {S.~N.}\ \bibnamefont
  {Santalla}},\ and\ \bibinfo {author} {\bibfnamefont {J.}~\bibnamefont
  {Rodr\'{\i}guez-Laguna}},\ }\href
  {https://doi.org/10.1103/PhysRevB.107.075116} {\bibfield  {journal} {\bibinfo
   {journal} {Phys. Rev. B}\ }\textbf {\bibinfo {volume} {107}},\ \bibinfo
  {pages} {075116} (\bibinfo {year} {2023})}\BibitemShut {NoStop}%
\bibitem [{\citenamefont {Ghosh}\ \emph {et~al.}(2020)\citenamefont {Ghosh},
  \citenamefont {Chanda},\ and\ \citenamefont {Sen(De)}}]{Ghosh_2020}%
  \BibitemOpen
  \bibfield  {author} {\bibinfo {author} {\bibfnamefont {S.}~\bibnamefont
  {Ghosh}}, \bibinfo {author} {\bibfnamefont {T.}~\bibnamefont {Chanda}},\ and\
  \bibinfo {author} {\bibfnamefont {A.}~\bibnamefont {Sen(De)}},\ }\href
  {https://doi.org/10.1103/PhysRevA.101.032115} {\bibfield  {journal} {\bibinfo
   {journal} {Phys. Rev. A}\ }\textbf {\bibinfo {volume} {101}},\ \bibinfo
  {pages} {032115} (\bibinfo {year} {2020})}\BibitemShut {NoStop}%
\bibitem [{\citenamefont {Zhao}\ \emph {et~al.}(2021)\citenamefont {Zhao},
  \citenamefont {Dou},\ and\ \citenamefont {Zhao}}]{Zhao_2021}%
  \BibitemOpen
  \bibfield  {author} {\bibinfo {author} {\bibfnamefont {F.}~\bibnamefont
  {Zhao}}, \bibinfo {author} {\bibfnamefont {F.-Q.}\ \bibnamefont {Dou}},\ and\
  \bibinfo {author} {\bibfnamefont {Q.}~\bibnamefont {Zhao}},\ }\href
  {https://doi.org/10.1103/PhysRevA.103.033715} {\bibfield  {journal} {\bibinfo
   {journal} {Phys. Rev. A}\ }\textbf {\bibinfo {volume} {103}},\ \bibinfo
  {pages} {033715} (\bibinfo {year} {2021})}\BibitemShut {NoStop}%
\bibitem [{\citenamefont {Rossini}\ \emph {et~al.}(2019)\citenamefont
  {Rossini}, \citenamefont {Andolina},\ and\ \citenamefont
  {Polini}}]{Rossini_2019}%
  \BibitemOpen
  \bibfield  {author} {\bibinfo {author} {\bibfnamefont {D.}~\bibnamefont
  {Rossini}}, \bibinfo {author} {\bibfnamefont {G.~M.}\ \bibnamefont
  {Andolina}},\ and\ \bibinfo {author} {\bibfnamefont {M.}~\bibnamefont
  {Polini}},\ }\href {https://doi.org/10.1103/PhysRevB.100.115142} {\bibfield
  {journal} {\bibinfo  {journal} {Phys. Rev. B}\ }\textbf {\bibinfo {volume}
  {100}},\ \bibinfo {pages} {115142} (\bibinfo {year} {2019})}\BibitemShut
  {NoStop}%
\bibitem [{\citenamefont {Ghosh}\ \emph {et~al.}(2021)\citenamefont {Ghosh},
  \citenamefont {Chanda}, \citenamefont {Mal},\ and\ \citenamefont
  {Sen(De)}}]{Ghosh_2021}%
  \BibitemOpen
  \bibfield  {author} {\bibinfo {author} {\bibfnamefont {S.}~\bibnamefont
  {Ghosh}}, \bibinfo {author} {\bibfnamefont {T.}~\bibnamefont {Chanda}},
  \bibinfo {author} {\bibfnamefont {S.}~\bibnamefont {Mal}},\ and\ \bibinfo
  {author} {\bibfnamefont {A.}~\bibnamefont {Sen(De)}},\ }\href
  {https://doi.org/10.1103/PhysRevA.104.032207} {\bibfield  {journal} {\bibinfo
   {journal} {Phys. Rev. A}\ }\textbf {\bibinfo {volume} {104}},\ \bibinfo
  {pages} {032207} (\bibinfo {year} {2021})}\BibitemShut {NoStop}%
\bibitem [{\citenamefont {Francica}\ \emph {et~al.}(2017)\citenamefont
  {Francica}, \citenamefont {Goold}, \citenamefont {Plastina},\ and\
  \citenamefont {Paternostro}}]{Francica_2017}%
  \BibitemOpen
  \bibfield  {author} {\bibinfo {author} {\bibfnamefont {G.}~\bibnamefont
  {Francica}}, \bibinfo {author} {\bibfnamefont {J.}~\bibnamefont {Goold}},
  \bibinfo {author} {\bibfnamefont {F.}~\bibnamefont {Plastina}},\ and\
  \bibinfo {author} {\bibfnamefont {M.}~\bibnamefont {Paternostro}},\ }\href
  {https://doi.org/10.1038/s41534-017-0012-8} {\bibfield  {journal} {\bibinfo
  {journal} {Npj Quantum Inf.}\ }\textbf {\bibinfo {volume} {3}},\ \bibinfo
  {pages} {s41534} (\bibinfo {year} {2017})}\BibitemShut {NoStop}%
\bibitem [{\citenamefont {Francica}\ \emph {et~al.}(2020)\citenamefont
  {Francica}, \citenamefont {Binder}, \citenamefont {Guarnieri}, \citenamefont
  {Mitchison}, \citenamefont {Goold},\ and\ \citenamefont
  {Plastina}}]{Francica_2020}%
  \BibitemOpen
  \bibfield  {author} {\bibinfo {author} {\bibfnamefont {G.}~\bibnamefont
  {Francica}}, \bibinfo {author} {\bibfnamefont {F.~C.}\ \bibnamefont
  {Binder}}, \bibinfo {author} {\bibfnamefont {G.}~\bibnamefont {Guarnieri}},
  \bibinfo {author} {\bibfnamefont {M.~T.}\ \bibnamefont {Mitchison}}, \bibinfo
  {author} {\bibfnamefont {J.}~\bibnamefont {Goold}},\ and\ \bibinfo {author}
  {\bibfnamefont {F.}~\bibnamefont {Plastina}},\ }\href
  {https://doi.org/10.1103/PhysRevLett.125.180603} {\bibfield  {journal}
  {\bibinfo  {journal} {Phys. Rev. Lett.}\ }\textbf {\bibinfo {volume} {125}},\
  \bibinfo {pages} {180603} (\bibinfo {year} {2020})}\BibitemShut {NoStop}%
\bibitem [{\citenamefont {Francica}(2024)}]{Francica_2024}%
  \BibitemOpen
  \bibfield  {author} {\bibinfo {author} {\bibfnamefont {G.}~\bibnamefont
  {Francica}},\ }\href {https://doi.org/10.1103/PhysRevA.110.062209} {\bibfield
   {journal} {\bibinfo  {journal} {Phys. Rev. A}\ }\textbf {\bibinfo {volume}
  {110}},\ \bibinfo {pages} {062209} (\bibinfo {year} {2024})}\BibitemShut
  {NoStop}%
\bibitem [{\citenamefont {Campaioli}\ \emph {et~al.}(2024)\citenamefont
  {Campaioli}, \citenamefont {Gherardini}, \citenamefont {Quach}, \citenamefont
  {Polini},\ and\ \citenamefont {Andolina}}]{Campaioli_2024}%
  \BibitemOpen
  \bibfield  {author} {\bibinfo {author} {\bibfnamefont {F.}~\bibnamefont
  {Campaioli}}, \bibinfo {author} {\bibfnamefont {S.}~\bibnamefont
  {Gherardini}}, \bibinfo {author} {\bibfnamefont {J.~Q.}\ \bibnamefont
  {Quach}}, \bibinfo {author} {\bibfnamefont {M.}~\bibnamefont {Polini}},\ and\
  \bibinfo {author} {\bibfnamefont {G.~M.}\ \bibnamefont {Andolina}},\ }\href
  {https://doi.org/10.1103/RevModPhys.96.031001} {\bibfield  {journal}
  {\bibinfo  {journal} {Rev. Mod. Phys.}\ }\textbf {\bibinfo {volume} {96}},\
  \bibinfo {pages} {031001} (\bibinfo {year} {2024})}\BibitemShut {NoStop}%
\bibitem [{\citenamefont {Perarnau-Llobet}\ \emph {et~al.}(2015)\citenamefont
  {Perarnau-Llobet}, \citenamefont {Hovhannisyan}, \citenamefont {Huber},
  \citenamefont {Skrzypczyk}, \citenamefont {Brunner},\ and\ \citenamefont
  {Ac\'{\i}n}}]{Perarnau_2015}%
  \BibitemOpen
  \bibfield  {author} {\bibinfo {author} {\bibfnamefont {M.}~\bibnamefont
  {Perarnau-Llobet}}, \bibinfo {author} {\bibfnamefont {K.~V.}\ \bibnamefont
  {Hovhannisyan}}, \bibinfo {author} {\bibfnamefont {M.}~\bibnamefont {Huber}},
  \bibinfo {author} {\bibfnamefont {P.}~\bibnamefont {Skrzypczyk}}, \bibinfo
  {author} {\bibfnamefont {N.}~\bibnamefont {Brunner}},\ and\ \bibinfo {author}
  {\bibfnamefont {A.}~\bibnamefont {Ac\'{\i}n}},\ }\href
  {https://doi.org/10.1103/PhysRevX.5.041011} {\bibfield  {journal} {\bibinfo
  {journal} {Phys. Rev. X}\ }\textbf {\bibinfo {volume} {5}},\ \bibinfo {pages}
  {041011} (\bibinfo {year} {2015})}\BibitemShut {NoStop}%
\bibitem [{\citenamefont {Rosa}\ \emph {et~al.}(2020)\citenamefont {Rosa},
  \citenamefont {Rossini}, \citenamefont {Andolina}, \citenamefont {Polini},\
  and\ \citenamefont {Carrega}}]{Rosa_2020}%
  \BibitemOpen
  \bibfield  {author} {\bibinfo {author} {\bibfnamefont {D.}~\bibnamefont
  {Rosa}}, \bibinfo {author} {\bibfnamefont {D.}~\bibnamefont {Rossini}},
  \bibinfo {author} {\bibfnamefont {G.~M.}\ \bibnamefont {Andolina}}, \bibinfo
  {author} {\bibfnamefont {M.}~\bibnamefont {Polini}},\ and\ \bibinfo {author}
  {\bibfnamefont {M.}~\bibnamefont {Carrega}},\ }\href
  {https://doi.org/10.1007/jhep11(2020)067} {\bibfield  {journal} {\bibinfo
  {journal} {J. High Energy Phys.}\ }\textbf {\bibinfo {volume} {2020}}\bibinfo
   {number} { (11)}}\BibitemShut {NoStop}%
\bibitem [{\citenamefont {Shaghaghi}\ \emph {et~al.}(2022)\citenamefont
  {Shaghaghi}, \citenamefont {Singh}, \citenamefont {Benenti},\ and\
  \citenamefont {Rosa}}]{Shaghaghi_2022}%
  \BibitemOpen
\bibfield  {number} {  }\bibfield  {author} {\bibinfo {author} {\bibfnamefont
  {V.}~\bibnamefont {Shaghaghi}}, \bibinfo {author} {\bibfnamefont
  {V.}~\bibnamefont {Singh}}, \bibinfo {author} {\bibfnamefont
  {G.}~\bibnamefont {Benenti}},\ and\ \bibinfo {author} {\bibfnamefont
  {D.}~\bibnamefont {Rosa}},\ }\href {https://doi.org/10.1088/2058-9565/ac8829}
  {\bibfield  {journal} {\bibinfo  {journal} {Quantum Sci. Technol.}\ }\textbf
  {\bibinfo {volume} {7}},\ \bibinfo {pages} {04LT01} (\bibinfo {year}
  {2022})}\BibitemShut {NoStop}%
\bibitem [{\citenamefont {Catalano}\ \emph {et~al.}(2023)\citenamefont
  {Catalano}, \citenamefont {Giampaolo}, \citenamefont {Morsch}, \citenamefont
  {Giovannetti},\ and\ \citenamefont {Franchini}}]{catalano_2023}%
  \BibitemOpen
  \bibfield  {author} {\bibinfo {author} {\bibfnamefont {A.~G.}\ \bibnamefont
  {Catalano}}, \bibinfo {author} {\bibfnamefont {S.~M.}\ \bibnamefont
  {Giampaolo}}, \bibinfo {author} {\bibfnamefont {O.}~\bibnamefont {Morsch}},
  \bibinfo {author} {\bibfnamefont {V.}~\bibnamefont {Giovannetti}},\ and\
  \bibinfo {author} {\bibfnamefont {F.}~\bibnamefont {Franchini}},\ }\href
  {https://arxiv.org/abs/2307.02529} {\bibinfo {title} {Frustrating quantum
  batteries}} (\bibinfo {year} {2023}),\ \Eprint
  {https://arxiv.org/abs/2307.02529} {arXiv:2307.02529 [quant-ph]} \BibitemShut
  {NoStop}%
\bibitem [{\citenamefont {Gyhm}\ \emph {et~al.}(2022)\citenamefont {Gyhm},
  \citenamefont {\ifmmode~\check{S}\else \v{S}\fi{}afr\'anek},\ and\
  \citenamefont {Rosa}}]{Gyhm_2022}%
  \BibitemOpen
  \bibfield  {author} {\bibinfo {author} {\bibfnamefont {J.-Y.}\ \bibnamefont
  {Gyhm}}, \bibinfo {author} {\bibfnamefont {D.}~\bibnamefont
  {\ifmmode~\check{S}\else \v{S}\fi{}afr\'anek}},\ and\ \bibinfo {author}
  {\bibfnamefont {D.}~\bibnamefont {Rosa}},\ }\href
  {https://doi.org/10.1103/PhysRevLett.128.140501} {\bibfield  {journal}
  {\bibinfo  {journal} {Phys. Rev. Lett.}\ }\textbf {\bibinfo {volume} {128}},\
  \bibinfo {pages} {140501} (\bibinfo {year} {2022})}\BibitemShut {NoStop}%
\bibitem [{\citenamefont {Gyhm}\ and\ \citenamefont
  {Fischer}(2024)}]{Gyhm_2024}%
  \BibitemOpen
  \bibfield  {author} {\bibinfo {author} {\bibfnamefont {J.-Y.}\ \bibnamefont
  {Gyhm}}\ and\ \bibinfo {author} {\bibfnamefont {U.~R.}\ \bibnamefont
  {Fischer}},\ }\href {https://doi.org/10.1116/5.0184903} {\bibfield  {journal}
  {\bibinfo  {journal} {AVS Quantum Sci.}\ }\textbf {\bibinfo {volume} {6}},\
  \bibinfo {pages} {012001} (\bibinfo {year} {2024})}\BibitemShut {NoStop}%
\bibitem [{\citenamefont {Yang}\ \emph {et~al.}(2025)\citenamefont {Yang},
  \citenamefont {Zhang}, \citenamefont {Wang},\ and\ \citenamefont
  {Shi}}]{Yang_2025}%
  \BibitemOpen
  \bibfield  {author} {\bibinfo {author} {\bibfnamefont {H.-Y.}\ \bibnamefont
  {Yang}}, \bibinfo {author} {\bibfnamefont {K.}~\bibnamefont {Zhang}},
  \bibinfo {author} {\bibfnamefont {X.-H.}\ \bibnamefont {Wang}},\ and\
  \bibinfo {author} {\bibfnamefont {H.-L.}\ \bibnamefont {Shi}},\ }\href
  {https://doi.org/10.1103/PhysRevB.111.085410} {\bibfield  {journal} {\bibinfo
   {journal} {Phys. Rev. B}\ }\textbf {\bibinfo {volume} {111}},\ \bibinfo
  {pages} {085410} (\bibinfo {year} {2025})}\BibitemShut {NoStop}%
\bibitem [{\citenamefont {Liu}\ \emph {et~al.}(2021)\citenamefont {Liu},
  \citenamefont {Shi}, \citenamefont {Shi}, \citenamefont {Wang},\ and\
  \citenamefont {Yang}}]{Liu_2021}%
  \BibitemOpen
  \bibfield  {author} {\bibinfo {author} {\bibfnamefont {J.-X.}\ \bibnamefont
  {Liu}}, \bibinfo {author} {\bibfnamefont {H.-L.}\ \bibnamefont {Shi}},
  \bibinfo {author} {\bibfnamefont {Y.-H.}\ \bibnamefont {Shi}}, \bibinfo
  {author} {\bibfnamefont {X.-H.}\ \bibnamefont {Wang}},\ and\ \bibinfo
  {author} {\bibfnamefont {W.-L.}\ \bibnamefont {Yang}},\ }\href
  {https://doi.org/10.1103/physrevb.104.245418} {\bibfield  {journal} {\bibinfo
   {journal} {Phys. Rev. B}\ }\textbf {\bibinfo {volume} {104}},\ \bibinfo
  {pages} {245418} (\bibinfo {year} {2021})}\BibitemShut {NoStop}%
\bibitem [{\citenamefont {Dou}\ \emph {et~al.}(2022)\citenamefont {Dou},
  \citenamefont {Zhou},\ and\ \citenamefont {Sun}}]{Dou_2022}%
  \BibitemOpen
  \bibfield  {author} {\bibinfo {author} {\bibfnamefont {F.-Q.}\ \bibnamefont
  {Dou}}, \bibinfo {author} {\bibfnamefont {H.}~\bibnamefont {Zhou}},\ and\
  \bibinfo {author} {\bibfnamefont {J.-A.}\ \bibnamefont {Sun}},\ }\href
  {https://doi.org/10.1103/PhysRevA.106.032212} {\bibfield  {journal} {\bibinfo
   {journal} {Phys. Rev. A}\ }\textbf {\bibinfo {volume} {106}},\ \bibinfo
  {pages} {032212} (\bibinfo {year} {2022})}\BibitemShut {NoStop}%
\bibitem [{\citenamefont {Andolina}\ \emph
  {et~al.}(2019{\natexlab{b}})\citenamefont {Andolina}, \citenamefont {Keck},
  \citenamefont {Mari}, \citenamefont {Giovannetti},\ and\ \citenamefont
  {Polini}}]{andolina_qm_class}%
  \BibitemOpen
  \bibfield  {author} {\bibinfo {author} {\bibfnamefont {G.~M.}\ \bibnamefont
  {Andolina}}, \bibinfo {author} {\bibfnamefont {M.}~\bibnamefont {Keck}},
  \bibinfo {author} {\bibfnamefont {A.}~\bibnamefont {Mari}}, \bibinfo {author}
  {\bibfnamefont {V.}~\bibnamefont {Giovannetti}},\ and\ \bibinfo {author}
  {\bibfnamefont {M.}~\bibnamefont {Polini}},\ }\href
  {https://doi.org/10.1103/PhysRevB.99.205437} {\bibfield  {journal} {\bibinfo
  {journal} {Phys. Rev. B}\ }\textbf {\bibinfo {volume} {99}},\ \bibinfo
  {pages} {205437} (\bibinfo {year} {2019}{\natexlab{b}})}\BibitemShut
  {NoStop}%
\bibitem [{\citenamefont {Gherardini}\ \emph
  {et~al.}(2020{\natexlab{a}})\citenamefont {Gherardini}, \citenamefont
  {Campaioli}, \citenamefont {Caruso},\ and\ \citenamefont {Binder}}]{gher}%
  \BibitemOpen
  \bibfield  {author} {\bibinfo {author} {\bibfnamefont {S.}~\bibnamefont
  {Gherardini}}, \bibinfo {author} {\bibfnamefont {F.}~\bibnamefont
  {Campaioli}}, \bibinfo {author} {\bibfnamefont {F.}~\bibnamefont {Caruso}},\
  and\ \bibinfo {author} {\bibfnamefont {F.~C.}\ \bibnamefont {Binder}},\
  }\href {https://doi.org/10.1103/PhysRevResearch.2.013095} {\bibfield
  {journal} {\bibinfo  {journal} {Phys. Rev. Res.}\ }\textbf {\bibinfo {volume}
  {2}},\ \bibinfo {pages} {013095} (\bibinfo {year}
  {2020}{\natexlab{a}})}\BibitemShut {NoStop}%
\bibitem [{\citenamefont {Gherardini}\ \emph
  {et~al.}(2020{\natexlab{b}})\citenamefont {Gherardini}, \citenamefont
  {Campaioli}, \citenamefont {Caruso},\ and\ \citenamefont
  {Binder}}]{Gherardini}%
  \BibitemOpen
  \bibfield  {author} {\bibinfo {author} {\bibfnamefont {S.}~\bibnamefont
  {Gherardini}}, \bibinfo {author} {\bibfnamefont {F.}~\bibnamefont
  {Campaioli}}, \bibinfo {author} {\bibfnamefont {F.}~\bibnamefont {Caruso}},\
  and\ \bibinfo {author} {\bibfnamefont {F.~C.}\ \bibnamefont {Binder}},\
  }\href {https://doi.org/10.1103/PhysRevResearch.2.013095} {\bibfield
  {journal} {\bibinfo  {journal} {Phys. Rev. Res.}\ }\textbf {\bibinfo {volume}
  {2}},\ \bibinfo {pages} {013095} (\bibinfo {year}
  {2020}{\natexlab{b}})}\BibitemShut {NoStop}%
\bibitem [{\citenamefont {Yao}\ and\ \citenamefont {Shao}(2021)}]{Yao_Rydberg}%
  \BibitemOpen
  \bibfield  {author} {\bibinfo {author} {\bibfnamefont {Y.}~\bibnamefont
  {Yao}}\ and\ \bibinfo {author} {\bibfnamefont {X.~Q.}\ \bibnamefont {Shao}},\
  }\href {https://doi.org/10.1103/PhysRevE.104.044116} {\bibfield  {journal}
  {\bibinfo  {journal} {Phys. Rev. E}\ }\textbf {\bibinfo {volume} {104}},\
  \bibinfo {pages} {044116} (\bibinfo {year} {2021})}\BibitemShut {NoStop}%
\bibitem [{\citenamefont {Grazi}\ \emph {et~al.}(2025)\citenamefont {Grazi},
  \citenamefont {Cavaliere}, \citenamefont {Sassetti}, \citenamefont
  {Ferraro},\ and\ \citenamefont {Traverso~Ziani}}]{Grazi}%
  \BibitemOpen
  \bibfield  {author} {\bibinfo {author} {\bibfnamefont {R.}~\bibnamefont
  {Grazi}}, \bibinfo {author} {\bibfnamefont {F.}~\bibnamefont {Cavaliere}},
  \bibinfo {author} {\bibfnamefont {M.}~\bibnamefont {Sassetti}}, \bibinfo
  {author} {\bibfnamefont {D.}~\bibnamefont {Ferraro}},\ and\ \bibinfo {author}
  {\bibfnamefont {N.}~\bibnamefont {Traverso~Ziani}},\ }\href
  {https://doi.org/10.1016/j.chaos.2025.116383} {\bibfield  {journal} {\bibinfo
   {journal} {Chaos Solit. Fractals}\ }\textbf {\bibinfo {volume} {196}},\
  \bibinfo {pages} {116383} (\bibinfo {year} {2025})}\BibitemShut {NoStop}%
\bibitem [{\citenamefont {Romero}\ \emph {et~al.}(2025)\citenamefont {Romero},
  \citenamefont {Chen},\ and\ \citenamefont {Ban}}]{romero_kickedisingQB}%
  \BibitemOpen
  \bibfield  {author} {\bibinfo {author} {\bibfnamefont {S.~V.}\ \bibnamefont
  {Romero}}, \bibinfo {author} {\bibfnamefont {X.}~\bibnamefont {Chen}},\ and\
  \bibinfo {author} {\bibfnamefont {Y.}~\bibnamefont {Ban}},\ }\href
  {https://arxiv.org/abs/2511.17835} {} (\bibinfo {year} {2025}),\ \Eprint
  {https://arxiv.org/abs/2511.17835} {arXiv:2511.17835 [quant-ph]} \BibitemShut
  {NoStop}%
\bibitem [{\citenamefont {Mazumdar}\ \emph {et~al.}(2025)\citenamefont
  {Mazumdar}, \citenamefont {Mitra},\ and\ \citenamefont
  {Srivastava}}]{mazumdar_mitra_SCL2025}%
  \BibitemOpen
  \bibfield  {author} {\bibinfo {author} {\bibfnamefont {A.}~\bibnamefont
  {Mazumdar}}, \bibinfo {author} {\bibfnamefont {A.}~\bibnamefont {Mitra}},\
  and\ \bibinfo {author} {\bibfnamefont {S.~C.~L.}\ \bibnamefont
  {Srivastava}},\ }\href {https://arxiv.org/abs/2511.22349} {} (\bibinfo {year}
  {2025}),\ \Eprint {https://arxiv.org/abs/2511.22349} {arXiv:2511.22349
  [quant-ph]} \BibitemShut {NoStop}%
\bibitem [{\citenamefont {Lu}\ \emph {et~al.}(2025)\citenamefont {Lu},
  \citenamefont {Tian}, \citenamefont {L\"u},\ and\ \citenamefont
  {Shang}}]{topological_QB}%
  \BibitemOpen
  \bibfield  {author} {\bibinfo {author} {\bibfnamefont {Z.-G.}\ \bibnamefont
  {Lu}}, \bibinfo {author} {\bibfnamefont {G.}~\bibnamefont {Tian}}, \bibinfo
  {author} {\bibfnamefont {X.-Y.}\ \bibnamefont {L\"u}},\ and\ \bibinfo
  {author} {\bibfnamefont {C.}~\bibnamefont {Shang}},\ }\href
  {https://doi.org/10.1103/PhysRevLett.134.180401} {\bibfield  {journal}
  {\bibinfo  {journal} {Phys. Rev. Lett.}\ }\textbf {\bibinfo {volume} {134}},\
  \bibinfo {pages} {180401} (\bibinfo {year} {2025})}\BibitemShut {NoStop}%
\bibitem [{\citenamefont {Rinaldi}\ \emph {et~al.}(2025)\citenamefont
  {Rinaldi}, \citenamefont {Filip}, \citenamefont {Gerace},\ and\ \citenamefont
  {Guarnieri}}]{rinaldi25}%
  \BibitemOpen
  \bibfield  {author} {\bibinfo {author} {\bibfnamefont {D.}~\bibnamefont
  {Rinaldi}}, \bibinfo {author} {\bibfnamefont {R.}~\bibnamefont {Filip}},
  \bibinfo {author} {\bibfnamefont {D.}~\bibnamefont {Gerace}},\ and\ \bibinfo
  {author} {\bibfnamefont {G.}~\bibnamefont {Guarnieri}},\ }\href
  {https://doi.org/10.1103/6kwv-z6fx} {\bibfield  {journal} {\bibinfo
  {journal} {Phys. Rev. A}\ }\textbf {\bibinfo {volume} {112}},\ \bibinfo
  {pages} {012205} (\bibinfo {year} {2025})}\BibitemShut {NoStop}%
\bibitem [{\citenamefont {Shukla}\ \emph {et~al.}(2025)\citenamefont {Shukla},
  \citenamefont {Kumar}, \citenamefont {Sen},\ and\ \citenamefont
  {Mishra}}]{shukla2025}%
  \BibitemOpen
  \bibfield  {author} {\bibinfo {author} {\bibfnamefont {R.~K.}\ \bibnamefont
  {Shukla}}, \bibinfo {author} {\bibfnamefont {R.}~\bibnamefont {Kumar}},
  \bibinfo {author} {\bibfnamefont {U.}~\bibnamefont {Sen}},\ and\ \bibinfo
  {author} {\bibfnamefont {S.~K.}\ \bibnamefont {Mishra}},\ }\href
  {https://arxiv.org/abs/2505.08029} {} (\bibinfo {year} {2025}),\ \Eprint
  {https://arxiv.org/abs/2505.08029} {arXiv:2505.08029 [quant-ph]} \BibitemShut
  {NoStop}%
\bibitem [{\citenamefont {Santos}\ \emph {et~al.}(2019)\citenamefont {Santos},
  \citenamefont {\ifmmode~\mbox{\c{C}}\else \c{C}\fi{}akmak}, \citenamefont
  {Campbell},\ and\ \citenamefont {Zinner}}]{Santos}%
  \BibitemOpen
  \bibfield  {author} {\bibinfo {author} {\bibfnamefont {A.~C.}\ \bibnamefont
  {Santos}}, \bibinfo {author} {\bibfnamefont {B.~i. e. i. f. m.~c.}\
  \bibnamefont {\ifmmode~\mbox{\c{C}}\else \c{C}\fi{}akmak}}, \bibinfo {author}
  {\bibfnamefont {S.}~\bibnamefont {Campbell}},\ and\ \bibinfo {author}
  {\bibfnamefont {N.~T.}\ \bibnamefont {Zinner}},\ }\href
  {https://doi.org/10.1103/PhysRevE.100.032107} {\bibfield  {journal} {\bibinfo
   {journal} {Phys. Rev. E}\ }\textbf {\bibinfo {volume} {100}},\ \bibinfo
  {pages} {032107} (\bibinfo {year} {2019})}\BibitemShut {NoStop}%
\bibitem [{\citenamefont {Allahverdyan}\ \emph {et~al.}(2004)\citenamefont
  {Allahverdyan}, \citenamefont {Balian},\ and\ \citenamefont
  {Nieuwenhuizen}}]{Allahverdyan_2004}%
  \BibitemOpen
  \bibfield  {author} {\bibinfo {author} {\bibfnamefont {A.~E.}\ \bibnamefont
  {Allahverdyan}}, \bibinfo {author} {\bibfnamefont {R.}~\bibnamefont
  {Balian}},\ and\ \bibinfo {author} {\bibfnamefont {T.~M.}\ \bibnamefont
  {Nieuwenhuizen}},\ }\href {https://doi.org/10.1209/epl/i2004-10101-2}
  {\bibfield  {journal} {\bibinfo  {journal} {Europhys. Lett.}\ }\textbf
  {\bibinfo {volume} {67}},\ \bibinfo {pages} {565} (\bibinfo {year}
  {2004})}\BibitemShut {NoStop}%
\bibitem [{\citenamefont {Wang}\ and\ \citenamefont {Dou}(2025)}]{wang25_erg}%
  \BibitemOpen
  \bibfield  {author} {\bibinfo {author} {\bibfnamefont {C.-J.}\ \bibnamefont
  {Wang}}\ and\ \bibinfo {author} {\bibfnamefont {F.-Q.}\ \bibnamefont {Dou}},\
  }\href {https://arxiv.org/abs/2512.21855} {} (\bibinfo {year} {2025}),\
  \Eprint {https://arxiv.org/abs/2512.21855} {arXiv:2512.21855 [quant-ph]}
  \BibitemShut {NoStop}%
\bibitem [{\citenamefont {Yang}\ \emph {et~al.}(2026)\citenamefont {Yang},
  \citenamefont {Wang}, \citenamefont {Maleki}, \citenamefont {Munro},
  \citenamefont {Agarwal},\ and\ \citenamefont {Scully}}]{yang26}%
  \BibitemOpen
  \bibfield  {author} {\bibinfo {author} {\bibfnamefont {F.}~\bibnamefont
  {Yang}}, \bibinfo {author} {\bibfnamefont {H.}~\bibnamefont {Wang}}, \bibinfo
  {author} {\bibfnamefont {Y.}~\bibnamefont {Maleki}}, \bibinfo {author}
  {\bibfnamefont {W.~J.}\ \bibnamefont {Munro}}, \bibinfo {author}
  {\bibfnamefont {G.~S.}\ \bibnamefont {Agarwal}},\ and\ \bibinfo {author}
  {\bibfnamefont {M.~O.}\ \bibnamefont {Scully}},\ }\href
  {https://arxiv.org/abs/2605.17700} {} (\bibinfo {year} {2026}),\ \Eprint
  {https://arxiv.org/abs/2605.17700} {arXiv:2605.17700 [quant-ph]} \BibitemShut
  {NoStop}%
\bibitem [{\citenamefont {Andolina}\ \emph
  {et~al.}(2019{\natexlab{c}})\citenamefont {Andolina}, \citenamefont {Keck},
  \citenamefont {Mari}, \citenamefont {Campisi}, \citenamefont {Giovannetti},\
  and\ \citenamefont {Polini}}]{andolina19}%
  \BibitemOpen
  \bibfield  {author} {\bibinfo {author} {\bibfnamefont {G.~M.}\ \bibnamefont
  {Andolina}}, \bibinfo {author} {\bibfnamefont {M.}~\bibnamefont {Keck}},
  \bibinfo {author} {\bibfnamefont {A.}~\bibnamefont {Mari}}, \bibinfo {author}
  {\bibfnamefont {M.}~\bibnamefont {Campisi}}, \bibinfo {author} {\bibfnamefont
  {V.}~\bibnamefont {Giovannetti}},\ and\ \bibinfo {author} {\bibfnamefont
  {M.}~\bibnamefont {Polini}},\ }\href
  {https://doi.org/10.1103/PhysRevLett.122.047702} {\bibfield  {journal}
  {\bibinfo  {journal} {Phys. Rev. Lett.}\ }\textbf {\bibinfo {volume} {122}},\
  \bibinfo {pages} {047702} (\bibinfo {year} {2019}{\natexlab{c}})}\BibitemShut
  {NoStop}%
\bibitem [{\citenamefont {Mitra}\ and\ \citenamefont
  {Srivastava}(2024)}]{MitraSrivastava_2024}%
  \BibitemOpen
  \bibfield  {author} {\bibinfo {author} {\bibfnamefont {A.}~\bibnamefont
  {Mitra}}\ and\ \bibinfo {author} {\bibfnamefont {S.~C.~L.}\ \bibnamefont
  {Srivastava}},\ }\href {https://doi.org/10.1103/PhysRevA.110.012227}
  {\bibfield  {journal} {\bibinfo  {journal} {Phys. Rev. A}\ }\textbf {\bibinfo
  {volume} {110}},\ \bibinfo {pages} {012227} (\bibinfo {year}
  {2024})}\BibitemShut {NoStop}%
\bibitem [{\citenamefont {Yang}\ \emph {et~al.}(2024)\citenamefont {Yang},
  \citenamefont {Shi}, \citenamefont {Wan}, \citenamefont {Zhang},
  \citenamefont {Wang},\ and\ \citenamefont {Yang}}]{Tavis_yang}%
  \BibitemOpen
  \bibfield  {author} {\bibinfo {author} {\bibfnamefont {H.-Y.}\ \bibnamefont
  {Yang}}, \bibinfo {author} {\bibfnamefont {H.-L.}\ \bibnamefont {Shi}},
  \bibinfo {author} {\bibfnamefont {Q.-K.}\ \bibnamefont {Wan}}, \bibinfo
  {author} {\bibfnamefont {K.}~\bibnamefont {Zhang}}, \bibinfo {author}
  {\bibfnamefont {X.-H.}\ \bibnamefont {Wang}},\ and\ \bibinfo {author}
  {\bibfnamefont {W.-L.}\ \bibnamefont {Yang}},\ }\href
  {https://doi.org/10.1103/PhysRevA.109.012204} {\bibfield  {journal} {\bibinfo
   {journal} {Phys. Rev. A}\ }\textbf {\bibinfo {volume} {109}},\ \bibinfo
  {pages} {012204} (\bibinfo {year} {2024})}\BibitemShut {NoStop}%
\bibitem [{\citenamefont {Ukhtary}\ \emph {et~al.}(2023)\citenamefont
  {Ukhtary}, \citenamefont {Nugraha}, \citenamefont {Cahaya}, \citenamefont
  {Rusydi},\ and\ \citenamefont {Majidi}}]{Ukhtary_23}%
  \BibitemOpen
  \bibfield  {author} {\bibinfo {author} {\bibfnamefont {M.~S.}\ \bibnamefont
  {Ukhtary}}, \bibinfo {author} {\bibfnamefont {A.~R.~T.}\ \bibnamefont
  {Nugraha}}, \bibinfo {author} {\bibfnamefont {A.~B.}\ \bibnamefont {Cahaya}},
  \bibinfo {author} {\bibfnamefont {A.}~\bibnamefont {Rusydi}},\ and\ \bibinfo
  {author} {\bibfnamefont {M.~A.}\ \bibnamefont {Majidi}},\ }\href
  {https://doi.org/10.1063/5.0156618} {\bibfield  {journal} {\bibinfo
  {journal} {Appl. Phys. Lett.}\ }\textbf {\bibinfo {volume} {123}},\ \bibinfo
  {pages} {5.0156618} (\bibinfo {year} {2023})}\BibitemShut {NoStop}%
\bibitem [{\citenamefont {Rodr\'{\i}guez}\ \emph
  {et~al.}(2023{\natexlab{a}})\citenamefont {Rodr\'{\i}guez}, \citenamefont
  {Ahmadi}, \citenamefont {Mazurek}, \citenamefont {Barzanjeh}, \citenamefont
  {Alicki},\ and\ \citenamefont {Horodecki}}]{Rodriguez_QHO}%
  \BibitemOpen
  \bibfield  {author} {\bibinfo {author} {\bibfnamefont {R.~R.}\ \bibnamefont
  {Rodr\'{\i}guez}}, \bibinfo {author} {\bibfnamefont {B.}~\bibnamefont
  {Ahmadi}}, \bibinfo {author} {\bibfnamefont {P.}~\bibnamefont {Mazurek}},
  \bibinfo {author} {\bibfnamefont {S.}~\bibnamefont {Barzanjeh}}, \bibinfo
  {author} {\bibfnamefont {R.}~\bibnamefont {Alicki}},\ and\ \bibinfo {author}
  {\bibfnamefont {P.}~\bibnamefont {Horodecki}},\ }\href
  {https://doi.org/10.1103/PhysRevA.107.042419} {\bibfield  {journal} {\bibinfo
   {journal} {Phys. Rev. A}\ }\textbf {\bibinfo {volume} {107}},\ \bibinfo
  {pages} {042419} (\bibinfo {year} {2023}{\natexlab{a}})}\BibitemShut
  {NoStop}%
\bibitem [{\citenamefont {Qu}\ \emph {et~al.}(2023)\citenamefont {Qu},
  \citenamefont {Zhan}, \citenamefont {Lin},\ and\ \citenamefont
  {Xue}}]{Peng_2023}%
  \BibitemOpen
  \bibfield  {author} {\bibinfo {author} {\bibfnamefont {D.}~\bibnamefont
  {Qu}}, \bibinfo {author} {\bibfnamefont {X.}~\bibnamefont {Zhan}}, \bibinfo
  {author} {\bibfnamefont {H.}~\bibnamefont {Lin}},\ and\ \bibinfo {author}
  {\bibfnamefont {P.}~\bibnamefont {Xue}},\ }\href
  {https://doi.org/10.1103/PhysRevB.108.L180301} {\bibfield  {journal}
  {\bibinfo  {journal} {Phys. Rev. B}\ }\textbf {\bibinfo {volume} {108}},\
  \bibinfo {pages} {L180301} (\bibinfo {year} {2023})}\BibitemShut {NoStop}%
\bibitem [{\citenamefont {Izrailev}(1990)}]{IZRAILEV_1990}%
  \BibitemOpen
  \bibfield  {author} {\bibinfo {author} {\bibfnamefont {F.~M.}\ \bibnamefont
  {Izrailev}},\ }\href
  {https://doi.org/https://doi.org/10.1016/0370-1573(90)90067-C} {\bibfield
  {journal} {\bibinfo  {journal} {Phys. Rep.}\ }\textbf {\bibinfo {volume}
  {196}},\ \bibinfo {pages} {299} (\bibinfo {year} {1990})}\BibitemShut
  {NoStop}%
\bibitem [{\citenamefont {Casati}\ \emph {et~al.}(1979)\citenamefont {Casati},
  \citenamefont {Chirikov}, \citenamefont {Izraelev},\ and\ \citenamefont
  {Ford}}]{res1}%
  \BibitemOpen
  \bibfield  {author} {\bibinfo {author} {\bibfnamefont {G.}~\bibnamefont
  {Casati}}, \bibinfo {author} {\bibfnamefont {B.~V.}\ \bibnamefont
  {Chirikov}}, \bibinfo {author} {\bibfnamefont {F.~M.}\ \bibnamefont
  {Izraelev}},\ and\ \bibinfo {author} {\bibfnamefont {J.}~\bibnamefont
  {Ford}},\ }in\ \href@noop {} {\emph {\bibinfo {booktitle} {Stochastic
  Behavior in Classical and Quantum Hamiltonian Systems}}},\ \bibinfo {editor}
  {edited by\ \bibinfo {editor} {\bibfnamefont {G.}~\bibnamefont {Casati}}\
  and\ \bibinfo {editor} {\bibfnamefont {J.}~\bibnamefont {Ford}}}\ (\bibinfo
  {publisher} {Springer Berlin Heidelberg},\ \bibinfo {address} {Berlin,
  Heidelberg},\ \bibinfo {year} {1979})\ pp.\ \bibinfo {pages}
  {334--352}\BibitemShut {NoStop}%
\bibitem [{\citenamefont {Paul}\ \emph {et~al.}(2016)\citenamefont {Paul},
  \citenamefont {Pal},\ and\ \citenamefont {Santhanam}}]{Sanku_2016}%
  \BibitemOpen
  \bibfield  {author} {\bibinfo {author} {\bibfnamefont {S.}~\bibnamefont
  {Paul}}, \bibinfo {author} {\bibfnamefont {H.}~\bibnamefont {Pal}},\ and\
  \bibinfo {author} {\bibfnamefont {M.~S.}\ \bibnamefont {Santhanam}},\ }\href
  {https://doi.org/10.1103/PhysRevE.93.060203} {\bibfield  {journal} {\bibinfo
  {journal} {Phys. Rev. E}\ }\textbf {\bibinfo {volume} {93}},\ \bibinfo
  {pages} {060203(R)} (\bibinfo {year} {2016})}\BibitemShut {NoStop}%
\bibitem [{\citenamefont {Moore}\ \emph {et~al.}(1995)\citenamefont {Moore},
  \citenamefont {Robinson}, \citenamefont {Bharucha}, \citenamefont
  {Sundaram},\ and\ \citenamefont {Raizen}}]{Moore_1995}%
  \BibitemOpen
  \bibfield  {author} {\bibinfo {author} {\bibfnamefont {F.~L.}\ \bibnamefont
  {Moore}}, \bibinfo {author} {\bibfnamefont {J.~C.}\ \bibnamefont {Robinson}},
  \bibinfo {author} {\bibfnamefont {C.~F.}\ \bibnamefont {Bharucha}}, \bibinfo
  {author} {\bibfnamefont {B.}~\bibnamefont {Sundaram}},\ and\ \bibinfo
  {author} {\bibfnamefont {M.~G.}\ \bibnamefont {Raizen}},\ }\href
  {https://doi.org/10.1103/PhysRevLett.75.4598} {\bibfield  {journal} {\bibinfo
   {journal} {Phys. Rev. Lett.}\ }\textbf {\bibinfo {volume} {75}},\ \bibinfo
  {pages} {4598} (\bibinfo {year} {1995})}\BibitemShut {NoStop}%
\bibitem [{\citenamefont {Paul}\ and\ \citenamefont
  {Santhanam}(2018)}]{Sanku_2017}%
  \BibitemOpen
  \bibfield  {author} {\bibinfo {author} {\bibfnamefont {S.}~\bibnamefont
  {Paul}}\ and\ \bibinfo {author} {\bibfnamefont {M.~S.}\ \bibnamefont
  {Santhanam}},\ }\href {https://doi.org/10.1103/PhysRevE.97.032217} {\bibfield
   {journal} {\bibinfo  {journal} {Phys. Rev. E}\ }\textbf {\bibinfo {volume}
  {97}},\ \bibinfo {pages} {032217} (\bibinfo {year} {2018})}\BibitemShut
  {NoStop}%
\bibitem [{\citenamefont {Wu}\ \emph {et~al.}(2009)\citenamefont {Wu},
  \citenamefont {Tonyushkin},\ and\ \citenamefont {Prentiss}}]{Mara_2009}%
  \BibitemOpen
  \bibfield  {author} {\bibinfo {author} {\bibfnamefont {S.}~\bibnamefont
  {Wu}}, \bibinfo {author} {\bibfnamefont {A.}~\bibnamefont {Tonyushkin}},\
  and\ \bibinfo {author} {\bibfnamefont {M.~G.}\ \bibnamefont {Prentiss}},\
  }\href {https://doi.org/10.1103/PhysRevLett.103.034101} {\bibfield  {journal}
  {\bibinfo  {journal} {Phys. Rev. Lett.}\ }\textbf {\bibinfo {volume} {103}},\
  \bibinfo {pages} {034101} (\bibinfo {year} {2009})}\BibitemShut {NoStop}%
\bibitem [{\citenamefont {Daszuta}\ and\ \citenamefont
  {Andersen}(2012)}]{Mikkel_2012}%
  \BibitemOpen
  \bibfield  {author} {\bibinfo {author} {\bibfnamefont {B.}~\bibnamefont
  {Daszuta}}\ and\ \bibinfo {author} {\bibfnamefont {M.~F.}\ \bibnamefont
  {Andersen}},\ }\href {https://doi.org/10.1103/PhysRevA.86.043604} {\bibfield
  {journal} {\bibinfo  {journal} {Phys. Rev. A}\ }\textbf {\bibinfo {volume}
  {86}},\ \bibinfo {pages} {043604} (\bibinfo {year} {2012})}\BibitemShut
  {NoStop}%
\bibitem [{\citenamefont {Lundh}\ and\ \citenamefont
  {Wallin}(2005)}]{Lundh_2005}%
  \BibitemOpen
  \bibfield  {author} {\bibinfo {author} {\bibfnamefont {E.}~\bibnamefont
  {Lundh}}\ and\ \bibinfo {author} {\bibfnamefont {M.}~\bibnamefont {Wallin}},\
  }\href {https://doi.org/10.1103/PhysRevLett.94.110603} {\bibfield  {journal}
  {\bibinfo  {journal} {Phys. Rev. Lett.}\ }\textbf {\bibinfo {volume} {94}},\
  \bibinfo {pages} {110603} (\bibinfo {year} {2005})}\BibitemShut {NoStop}%
\bibitem [{\citenamefont {Dana}\ \emph {et~al.}(2008)\citenamefont {Dana},
  \citenamefont {Ramareddy}, \citenamefont {Talukdar},\ and\ \citenamefont
  {Summy}}]{Dana_2008}%
  \BibitemOpen
  \bibfield  {author} {\bibinfo {author} {\bibfnamefont {I.}~\bibnamefont
  {Dana}}, \bibinfo {author} {\bibfnamefont {V.}~\bibnamefont {Ramareddy}},
  \bibinfo {author} {\bibfnamefont {I.}~\bibnamefont {Talukdar}},\ and\
  \bibinfo {author} {\bibfnamefont {G.~S.}\ \bibnamefont {Summy}},\ }\href
  {https://doi.org/10.1103/PhysRevLett.100.024103} {\bibfield  {journal}
  {\bibinfo  {journal} {Phys. Rev. Lett.}\ }\textbf {\bibinfo {volume} {100}},\
  \bibinfo {pages} {024103} (\bibinfo {year} {2008})}\BibitemShut {NoStop}%
\bibitem [{\citenamefont {Delvecchio}\ \emph {et~al.}(2020)\citenamefont
  {Delvecchio}, \citenamefont {Petiziol},\ and\ \citenamefont
  {Wimberger}}]{Sandro_2020}%
  \BibitemOpen
  \bibfield  {author} {\bibinfo {author} {\bibfnamefont {M.}~\bibnamefont
  {Delvecchio}}, \bibinfo {author} {\bibfnamefont {F.}~\bibnamefont
  {Petiziol}},\ and\ \bibinfo {author} {\bibfnamefont {S.}~\bibnamefont
  {Wimberger}},\ }\bibfield  {journal} {\bibinfo  {journal} {Condensed Matter}\
  }\textbf {\bibinfo {volume} {5}},\ \href
  {https://doi.org/10.3390/condmat5010004} {10.3390/condmat5010004} (\bibinfo
  {year} {2020})\BibitemShut {NoStop}%
\bibitem [{\citenamefont {Paul}\ \emph {et~al.}(2024)\citenamefont {Paul},
  \citenamefont {Kannan},\ and\ \citenamefont {Santhanam}}]{S_paul_res}%
  \BibitemOpen
  \bibfield  {author} {\bibinfo {author} {\bibfnamefont {S.}~\bibnamefont
  {Paul}}, \bibinfo {author} {\bibfnamefont {J.~B.}\ \bibnamefont {Kannan}},\
  and\ \bibinfo {author} {\bibfnamefont {M.~S.}\ \bibnamefont {Santhanam}},\
  }\href {https://doi.org/10.1103/PhysRevB.110.144301} {\bibfield  {journal}
  {\bibinfo  {journal} {Phys. Rev. B}\ }\textbf {\bibinfo {volume} {110}},\
  \bibinfo {pages} {144301} (\bibinfo {year} {2024})}\BibitemShut {NoStop}%
\bibitem [{\citenamefont {Zhou}\ and\ \citenamefont {Wang}(2026)}]{Zhou_2026}%
  \BibitemOpen
  \bibfield  {author} {\bibinfo {author} {\bibfnamefont {Y.}~\bibnamefont
  {Zhou}}\ and\ \bibinfo {author} {\bibfnamefont {J.}~\bibnamefont {Wang}},\
  }\href {https://doi.org/10.1103/1lj8-xjps} {\bibfield  {journal} {\bibinfo
  {journal} {Phys. Rev. B}\ }\textbf {\bibinfo {volume} {113}},\ \bibinfo
  {pages} {144307} (\bibinfo {year} {2026})}\BibitemShut {NoStop}%
\bibitem [{\citenamefont {Haake}\ and\ \citenamefont
  {Shepelyansky}(1988)}]{Haake_1988}%
  \BibitemOpen
  \bibfield  {author} {\bibinfo {author} {\bibfnamefont {F.}~\bibnamefont
  {Haake}}\ and\ \bibinfo {author} {\bibfnamefont {D.~L.}\ \bibnamefont
  {Shepelyansky}},\ }\href {https://doi.org/10.1209/0295-5075/5/8/001}
  {\bibfield  {journal} {\bibinfo  {journal} {Europhysics Letters}\ }\textbf
  {\bibinfo {volume} {5}},\ \bibinfo {pages} {671} (\bibinfo {year}
  {1988})}\BibitemShut {NoStop}%
\bibitem [{\citenamefont {Chaudhury}\ \emph {et~al.}(2009)\citenamefont
  {Chaudhury}, \citenamefont {Smith}, \citenamefont {Anderson}, \citenamefont
  {Ghose},\ and\ \citenamefont {Jessen}}]{Jessen_2009}%
  \BibitemOpen
  \bibfield  {author} {\bibinfo {author} {\bibfnamefont {S.}~\bibnamefont
  {Chaudhury}}, \bibinfo {author} {\bibfnamefont {A.}~\bibnamefont {Smith}},
  \bibinfo {author} {\bibfnamefont {B.~E.}\ \bibnamefont {Anderson}}, \bibinfo
  {author} {\bibfnamefont {S.}~\bibnamefont {Ghose}},\ and\ \bibinfo {author}
  {\bibfnamefont {P.~S.}\ \bibnamefont {Jessen}},\ }\href
  {https://doi.org/10.1038/nature08396} {\bibfield  {journal} {\bibinfo
  {journal} {Nature}\ }\textbf {\bibinfo {volume} {461}},\ \bibinfo {pages}
  {768} (\bibinfo {year} {2009})}\BibitemShut {NoStop}%
\bibitem [{\citenamefont {Neill}\ \emph {et~al.}(2016)\citenamefont {Neill},
  \citenamefont {Roushan}, \citenamefont {Fang}, \citenamefont {Chen},
  \citenamefont {Kolodrubetz}, \citenamefont {Chen}, \citenamefont {Megrant},
  \citenamefont {Barends}, \citenamefont {Campbell}, \citenamefont {Chiaro},
  \citenamefont {Dunsworth}, \citenamefont {Jeffrey}, \citenamefont {Kelly},
  \citenamefont {Mutus}, \citenamefont {O’Malley}, \citenamefont {Quintana},
  \citenamefont {Sank}, \citenamefont {Vainsencher}, \citenamefont {Wenner},
  \citenamefont {White}, \citenamefont {Polkovnikov},\ and\ \citenamefont
  {Martinis}}]{Neill_2016}%
  \BibitemOpen
  \bibfield  {author} {\bibinfo {author} {\bibfnamefont {C.}~\bibnamefont
  {Neill}}, \bibinfo {author} {\bibfnamefont {P.}~\bibnamefont {Roushan}},
  \bibinfo {author} {\bibfnamefont {M.}~\bibnamefont {Fang}}, \bibinfo {author}
  {\bibfnamefont {Y.}~\bibnamefont {Chen}}, \bibinfo {author} {\bibfnamefont
  {M.}~\bibnamefont {Kolodrubetz}}, \bibinfo {author} {\bibfnamefont
  {Z.}~\bibnamefont {Chen}}, \bibinfo {author} {\bibfnamefont {A.}~\bibnamefont
  {Megrant}}, \bibinfo {author} {\bibfnamefont {R.}~\bibnamefont {Barends}},
  \bibinfo {author} {\bibfnamefont {B.}~\bibnamefont {Campbell}}, \bibinfo
  {author} {\bibfnamefont {B.}~\bibnamefont {Chiaro}}, \bibinfo {author}
  {\bibfnamefont {A.}~\bibnamefont {Dunsworth}}, \bibinfo {author}
  {\bibfnamefont {E.}~\bibnamefont {Jeffrey}}, \bibinfo {author} {\bibfnamefont
  {J.}~\bibnamefont {Kelly}}, \bibinfo {author} {\bibfnamefont
  {J.}~\bibnamefont {Mutus}}, \bibinfo {author} {\bibfnamefont {P.~J.~J.}\
  \bibnamefont {O’Malley}}, \bibinfo {author} {\bibfnamefont
  {C.}~\bibnamefont {Quintana}}, \bibinfo {author} {\bibfnamefont
  {D.}~\bibnamefont {Sank}}, \bibinfo {author} {\bibfnamefont {A.}~\bibnamefont
  {Vainsencher}}, \bibinfo {author} {\bibfnamefont {J.}~\bibnamefont {Wenner}},
  \bibinfo {author} {\bibfnamefont {T.~C.}\ \bibnamefont {White}}, \bibinfo
  {author} {\bibfnamefont {A.}~\bibnamefont {Polkovnikov}},\ and\ \bibinfo
  {author} {\bibfnamefont {J.~M.}\ \bibnamefont {Martinis}},\ }\href
  {https://doi.org/10.1038/nphys3830} {\bibfield  {journal} {\bibinfo
  {journal} {Nat. Phys.}\ }\textbf {\bibinfo {volume} {12}},\ \bibinfo {pages}
  {1037} (\bibinfo {year} {2016})}\BibitemShut {NoStop}%
\bibitem [{\citenamefont {Krithika}\ \emph {et~al.}(2019)\citenamefont
  {Krithika}, \citenamefont {Anjusha}, \citenamefont {Bhosale},\ and\
  \citenamefont {Mahesh}}]{Krithika_2019}%
  \BibitemOpen
  \bibfield  {author} {\bibinfo {author} {\bibfnamefont {V.~R.}\ \bibnamefont
  {Krithika}}, \bibinfo {author} {\bibfnamefont {V.~S.}\ \bibnamefont
  {Anjusha}}, \bibinfo {author} {\bibfnamefont {U.~T.}\ \bibnamefont
  {Bhosale}},\ and\ \bibinfo {author} {\bibfnamefont {T.~S.}\ \bibnamefont
  {Mahesh}},\ }\href {https://doi.org/10.1103/PhysRevE.99.032219} {\bibfield
  {journal} {\bibinfo  {journal} {Phys. Rev. E}\ }\textbf {\bibinfo {volume}
  {99}},\ \bibinfo {pages} {032219} (\bibinfo {year} {2019})}\BibitemShut
  {NoStop}%
\bibitem [{\citenamefont {Norcia}\ \emph {et~al.}(2018)\citenamefont {Norcia},
  \citenamefont {Lewis-Swan}, \citenamefont {Cline}, \citenamefont {Zhu},
  \citenamefont {Rey},\ and\ \citenamefont {Thompson}}]{Thompson_2018}%
  \BibitemOpen
  \bibfield  {author} {\bibinfo {author} {\bibfnamefont {M.~A.}\ \bibnamefont
  {Norcia}}, \bibinfo {author} {\bibfnamefont {R.~J.}\ \bibnamefont
  {Lewis-Swan}}, \bibinfo {author} {\bibfnamefont {J.~R.~K.}\ \bibnamefont
  {Cline}}, \bibinfo {author} {\bibfnamefont {B.}~\bibnamefont {Zhu}}, \bibinfo
  {author} {\bibfnamefont {A.~M.}\ \bibnamefont {Rey}},\ and\ \bibinfo {author}
  {\bibfnamefont {J.~K.}\ \bibnamefont {Thompson}},\ }\href
  {https://doi.org/10.1126/science.aar3102} {\bibfield  {journal} {\bibinfo
  {journal} {Science}\ }\textbf {\bibinfo {volume} {361}},\ \bibinfo {pages}
  {259} (\bibinfo {year} {2018})}\BibitemShut {NoStop}%
\bibitem [{\citenamefont {Rodr\'{\i}guez}\ \emph
  {et~al.}(2023{\natexlab{b}})\citenamefont {Rodr\'{\i}guez}, \citenamefont
  {Ahmadi}, \citenamefont {Mazurek}, \citenamefont {Barzanjeh}, \citenamefont
  {Alicki},\ and\ \citenamefont {Horodecki}}]{catalyst_QHO}%
  \BibitemOpen
  \bibfield  {author} {\bibinfo {author} {\bibfnamefont {R.~R.}\ \bibnamefont
  {Rodr\'{\i}guez}}, \bibinfo {author} {\bibfnamefont {B.}~\bibnamefont
  {Ahmadi}}, \bibinfo {author} {\bibfnamefont {P.}~\bibnamefont {Mazurek}},
  \bibinfo {author} {\bibfnamefont {S.}~\bibnamefont {Barzanjeh}}, \bibinfo
  {author} {\bibfnamefont {R.}~\bibnamefont {Alicki}},\ and\ \bibinfo {author}
  {\bibfnamefont {P.}~\bibnamefont {Horodecki}},\ }\href
  {https://doi.org/10.1103/PhysRevA.107.042419} {\bibfield  {journal} {\bibinfo
   {journal} {Phys. Rev. A}\ }\textbf {\bibinfo {volume} {107}},\ \bibinfo
  {pages} {042419} (\bibinfo {year} {2023}{\natexlab{b}})}\BibitemShut
  {NoStop}%
\bibitem [{\citenamefont {Andolina}\ \emph {et~al.}(2025)\citenamefont
  {Andolina}, \citenamefont {Stanzione}, \citenamefont {Giovannetti},\ and\
  \citenamefont {Polini}}]{andolina_25}%
  \BibitemOpen
  \bibfield  {author} {\bibinfo {author} {\bibfnamefont {G.~M.}\ \bibnamefont
  {Andolina}}, \bibinfo {author} {\bibfnamefont {V.}~\bibnamefont {Stanzione}},
  \bibinfo {author} {\bibfnamefont {V.}~\bibnamefont {Giovannetti}},\ and\
  \bibinfo {author} {\bibfnamefont {M.}~\bibnamefont {Polini}},\ }\href
  {https://doi.org/10.1103/kzvn-dj7v} {\bibfield  {journal} {\bibinfo
  {journal} {Phys. Rev. Lett.}\ }\textbf {\bibinfo {volume} {134}},\ \bibinfo
  {pages} {240403} (\bibinfo {year} {2025})}\BibitemShut {NoStop}%
\end{thebibliography}%


\onecolumngrid

\clearpage


\setcounter{equation}{0}
\setcounter{figure}{0}
\setcounter{table}{0}
\setcounter{page}{1}
\setcounter{section}{0} 

\renewcommand{\theequation}{S\arabic{equation}}
\renewcommand{\thefigure}{S\arabic{figure}}
\renewcommand{\thetable}{S\arabic{table}}
\renewcommand{\thesection}{S\Roman{section}}


\renewcommand{\bibnumfmt}[1]{[S#1]} 
\renewcommand{\citenumfont}[1]{S#1} 

\begin{center}
	\textbf{\large Supplemental Material for ``Quantum 
	resonance-enhanced performance of quantum battery"}
\end{center}
\vspace{10pt}
In this Supplemental Material, we provide additional details and calculations 
for the results stated in the main text. In Sec.~\ref{sec:E_analytics}, we 
have shown the analytical calculation of the energy growth for $K_i =0, K\ne 
0$. Guided by the numerics we also have modified the expression for $K_i \ne 
0$.  Next in Sec. \ref{sec:eff_k2} we have shown the variation of efficiency 
($\eta$) with respect to the kick strength ($K_2$). In Sec.~\ref{sec:SL_top} 
the dynamics of the linear entropy has been shown for interacting kicked top 
model under the resonance condition.

\refstepcounter{section} 
\section*{S\Roman{section}: Analytical expression for the stored energy 
within the rotor subsystem}\label{sec:E_analytics}
{Quantum resonance is a purely quantum phenomenon, entirely independent of 
the underlying classical dynamics. It emerges under the specific condition 
$\hbar_s T = 4\pi l/l^{\prime}$, where $l, l^{\prime} \in \mathbb{Z}$. 
Notably, this condition does not depend on the kick strengths of the 
individual rotors. Therefore, to derive an analytical expression for the 
stored energy in the presence of interaction, we set $K_1 = K_2 = 0$ and $K 
\neq 0$, which yields the Hamiltonian,}
\begin{equation}
	H= \frac{p_{1}^{2}}{2\mu_1}+ \frac{p_{2}^{2}}{2\mu_2}+ K 
	\cos(x_1-x_2)\sum_\tau \delta \left(t- \tau T \right).
\end{equation}
where $x_{1(2)}$ and $p_{1(2)}$ are the position and the momentum of the 
rotor $1(2)$, the mass of the rotor is $\mu_{1(2)}$. Here, we consider 
$\mu_{1(2)}^{-1} \in \mathbb{Z}$ and {$\mu_1\neq \mu_2$}. {Under the 
transformation, }
\begin{align}
	\Theta_1 = x_1 +x_2,\hspace{3mm} \Theta_2= x_1- x_2, \hspace{3mm} u = 
	\frac{p_1 +p_2}{2},\hspace{3mm} v= \frac{p_1 -p_2}{2},
\end{align}
{the Hamiltonian transforms to,}
\begin{align*}
	\Tilde{H} =\left[(u^2 +v^2)\left(\frac{\mu_1 +\mu_2}{2\mu_1 \mu_2}\right) 
	+2uv\left(\frac{\mu_2 -\mu_1}{2\mu_1\mu_2}\right) \right] + K\cos\Theta_2 
	\sum_r \delta \left(\tau- rT \right).
\end{align*}
The time evolution operator is given by,
\begin{equation}
	\begin{aligned}
		U&= U^{\text{free}} U^{\text{int}} =\left( U_{p_1}\otimes 
		U_{p_2}\right)U^{\text{int}}\\
		&= \mathcal{F}(u, v, \mu_1, \mu_2, T)\cdot e^{\frac{-iK\cos \Theta_2 
		}{\hbar_s}}\,,
	\end{aligned}     
\end{equation}
where $\hbar_s$ is the scaled Planck constant. The interaction manifests in 
momentum space as a complex function, $\mathcal{F}(u, v, \mu_1, \mu_2, T)$. 
Now imposing the condition for primary resonance $\hbar_s T= 4\pi$, we get,
\begin{equation}
	U =  \unity  e^{\frac{-iK\cos \Theta_2 }{\hbar_s}}
\end{equation}
Thus, the interaction term $U^\text{int}$ effectively reduces to a 
single-particle kick term. For this single-particle kicked rotor system, we 
consider the initial state as $\ket{\psi (0)}= \ket{n=0}$, where 
$\mathcal{P}\ket{n}= n\hbar_s \ket{n}$, $\mathcal{P}$ being the momentum of 
the particle.

The time evolved state {$\vert{} \psi(\tau)\rangle$ is then effectively 
obtained by time-evolution due to the ``single" particle} quantum kicked 
rotor (QKR) and is given by $\ket{\psi(\tau)}=\sum_{n}(-i)^n 
J_n\left(\frac{K\tau}{\hbar_S} \right)\ket{n}=\sum_n c_n(\tau)\ket{n}$, where 
$J_n(\cdot)$ is the Bessel function of first kind of order $n$. We know from 
Ref. \cite{S_paul_res} that the purity of the coupled QKR is equivalent to 
the participation ratio of this coordinate-transformed single-particle QKR. 
We have the single-particle density matrix as
\begin{align}
	\rho(\tau)&= 
	\ket{\psi(\tau)}\bra{\psi(\tau)}=\sum_{n,n^\prime}c_n(\tau)c_{n^\prime}(\tau)\ket{n}\bra{n^\prime}
\end{align}
By considering $\mu_2=1$, the energy stored within the rotor is given by,
\begin{align}
	\langle H_0\rangle&= \langle \sum_{n^{\prime \prime}} \frac{n^{\prime 
	\prime 2}\hbar_s^2}{2}\ket{n^{\prime \prime}}\bra{n^{\prime 
	\prime}}\rangle = \Tr\left(\sum_{n^{\prime \prime},n^\prime, 
	n}\frac{n^{\prime \prime 
	2}\hbar_s^2}{2}c_n(\tau)c_{n^\prime}(\tau)\ket{n^{\prime \prime}}\langle 
	n^{\prime \prime}|n\rangle \bra{n^\prime} \right) =\sum_{n, 
	n^\prime}\frac{n^2\hbar_s^2}{2}c_n(\tau)c_{n^\prime}(\tau) \langle 
	n|n\rangle \langle n^{\prime }|n\rangle,.
\end{align}
Using $\langle n^{\prime \prime}|n\rangle= \delta_{n^{\prime \prime},n}$, the 
stored energy simplified to,
\begin{equation}\label{eq:energy_supp}
	\begin{aligned}
		\langle H_0\rangle
		&= \sum_n \frac{n^2\hbar_s^2}{2}\lvert c_n(\tau) \rvert^2 = \sum_{n= 
		-\frac{L}{2}}^{\frac{L}{2}}\frac{n^2\hbar_s^2}{2}\left[J_n\left(\frac{K\tau}{\hbar_s}
		 \right) \right]^2,.
	\end{aligned}
\end{equation}
To simplify the above equation, we use two recurrence relations of the Bessel 
function of the first kind,
\begin{align}
	J_{n+1}(x)&= \frac{2n}{x}J_n(x) -J_{n-1}(x) \\
	\implies nJ_n(x)&= \left( J_{n-1}(x) +J_{n+1}(x)\right)\frac{x}{2}\,,
\end{align}
and $J_{-n}(x)= (-1)^n J_n(x)$, which leads to
\begin{align*}
	\langle H_0\rangle &= \sum_{n= 
		-\frac{L}{2}}^{\frac{L}{2}}\frac{n^2\hbar_s^2}{2}\left[J_n\left(\frac{K\tau}{\hbar_s}
	\right) \right]^2
	=\sum_{n=1}^{\frac{L}{2}}\hbar_s^2\left[n.J_n\left(\frac{K\tau}{\hbar_s} 
	\right) \right]^2
	= \sum_{n=1}^{\frac{L}{2}} 
	\left[J_{n-1}\left(\frac{K\tau}{\hbar_s}\right)+ 
	J_{n+1}\left(\frac{K\tau}{\hbar_s}\right) \right]^2 \hbar_s^2 \cdot 
	\left(\frac{K\tau}{2\hbar_s}\right)^2\\
	&=\left(\frac{K^2\tau^2}{4}\right) \cdot \sum_{n=1}^{\frac{L}{2}} 
	\left[J_{n-1}\left(\frac{K\tau}{\hbar_s}\right)+ 
	J_{n+1}\left(\frac{K\tau}{\hbar_s}\right) \right]^2\,.
\end{align*}

\begin{figure}[t]
	\centering
	\includegraphics{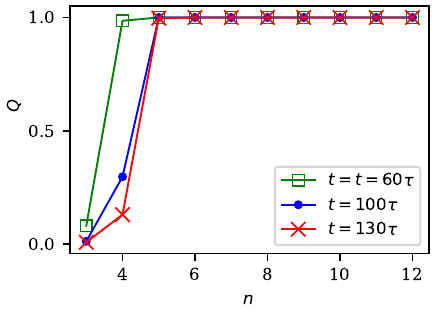}
	\caption{$Q$ has been plotted with $n$, where the system size is 
	considered as $L=2^n$ for three different times. The other parameters are 
	$K_1= K_2=0$, $K=0.1$, $\mu_1=2$, $\mu_2=1$ and $T=12$ keeping $\hbar_s 
	T=4\pi$. This can also be shown for $K_2 \ne0$.}
	\label{fig:sys}
\end{figure}
From Fig.~\ref{fig:sys} it can be observed that the quantity $Q= 
\sum_{n=1}^{\frac{L}{2}} \left[J_{n-1}\left(\frac{K\tau}{\hbar_s}\right)+ 
J_{n+1}\left(\frac{K\tau}{\hbar_s}\right) \right]^2$ becomes constant in time 
with increasing system size. Thus, the stored energy is approximately given 
by $\left(\frac{K^2\tau^2}{4}\right)$.

\subsection{Expression for the stored energy when $K_2\ne 0$}

\begin{figure}
	\centering
	\includegraphics{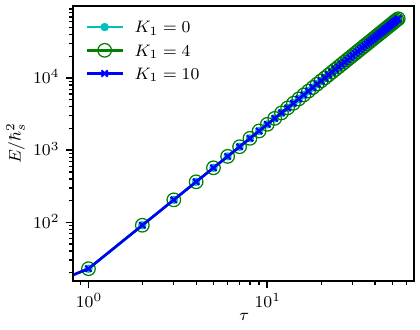}
	\caption{The dynamics of the stored energy is plotted for $K=0.1$ and 
	$K_2=10$ under the resonance condition i.e $\hbar_sT=4\pi$. The other 
	parameters are fixed as $\mu_1=0.5, \mu_2=1$. The independence of 
	$E({\tau})$ over $K_1$ occurs only for $\hbar_sT=4\pi$. }
	\label{fig:E_K1}
\end{figure}

From Figure~\ref{fig:E_K1}, it is evident that the stored energy of one of 
the rotors with kick strength $K_j$ does not depend on the kick strength of 
the other rotor. Thus, the energy stored within the rotor subsystem for $L 
\to \infty$ is given by the ansatz,
\begin{equation}\label{eq:ansatz}
	E(K, K_2, \tau)= (\alpha K^2 + \beta K_2^2 +\gamma K K_2 +\epsilon(K+ 
	K_2)+c )\tau^2 
\end{equation}
Now for $K_2=0$, the energy expression must take the form of 
Eq.~\ref{eq:energy_supp}, so we have $\alpha= 1/4$. Similarly, when $K=0$, 
the rotors become non-interacting, thus again $\beta= 1/4$ (expression of a 
single kicked rotor). So if we terminate the system size $L$ to some large 
but finite value, the expression of the stored energy is given by,
\begin{equation}\label{eq:E_kk2}
	E(K, K_2, \tau)= \sum_{n= -\frac{L}{2}}^{\frac{L}{2}} 
	\frac{n^2\hbar_s^2}{2}\left[J_n\left(\frac{K\tau}{\hbar_s} \right) 
	\right]^2 + 
	\frac{n^2\hbar_s^2}{2}\left[J_n\left(\frac{K_2\tau}{\hbar_s} \right) 
	\right]^2 
\end{equation}
Considering the above expression Eq.~\ref{eq:E_kk2} and matching it with 
numerical data, we have found that the third, fourth, and fifth terms in Eq. 
~\ref {eq:ansatz} do not contribute to the energy. Thus, we can set 
$\gamma=\epsilon=c=0$.

\refstepcounter{section} 
\section{S\Roman{section}: Variation of the efficiency with 
$K_2$}\label{sec:eff_k2}
At resonance, the passive state energy $\tilde{E}(\tau)$ is independent of 
$K_i$ and depends only on the interaction strength $K$. It was established in 
Ref.~\cite{S_paul_res} that entanglement growth is solely determined by $K$. 
Because the passive state energy quantifies the energy rendered inaccessible 
due to the mixedness of the system, this independence of $\tilde{E}(\tau)$ 
from $K_i$ is physically expected. Consequently, we obtain
\begin{equation}
	\eta(\tau)=
	\frac{
		\sum\limits_{n=-L/2}^{L/2}
		\left[
		J_n^2\!\left(\frac{K_2\tau}{\hbar_s}\right)
		+J_n^2\!\left(\frac{K\tau}{\hbar_s}\right)
		-\tilde{\lambda}_n^2(\tau)
		\right]
		\dfrac{n^2\hbar_s^2}{2}
	}{
		\sum\limits_{n=-L/2}^{L/2}
		\left[
		J_n^2\!\left(\frac{K_2\tau}{\hbar_s}\right)
		+J_n^2\!\left(\frac{K\tau}{\hbar_s}\right)
		\right]
		\dfrac{n^2\hbar_s^2}{2}
	}.
	\label{eq:efficiency}
\end{equation} 

\begin{figure}[htbt]
	\centering
	\includegraphics{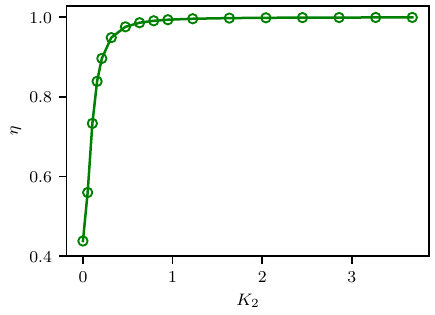}
	\caption{Variation of $\eta$ (at $\tau =50$) with $K_2$ is plotted for 
	$K= 0.1$ at the primary resonance condition. The remaining parameters are 
	fixed at: $K_1=9$, $\mu_1=2$, $\mu_2=1$, $L=2^{11}$ and $T=12$.}
	\label{fig:eff_k2}
\end{figure}

Consequently, the efficiency approaches unity ($\eta \sim 1$) in the limit 
$K_i \gg K$. By analyzing the dependence of $\eta$ on $K_2$, as depicted in 
Fig.~\ref{fig:eff_k2}, we observe that almost unit efficiency is successfully 
maintained for all $K_2 \geq 1$.

\refstepcounter{section} 
\section{S\Roman{section}: Entanglement growth of coupled kicked 
top}\label{sec:SL_top}
The period-$1$ time evolution operator ($\hbar=1$) for the coupled kicked top 
is given by
\begin{equation}
	\begin{aligned}
		U &= \exp\left( -i\sum_m  \frac{\beta_m}{2j} (J_{m}^{z})^2\right) 
		\exp(-i V_{\rm int}), 
		\quad  \text{where} \quad V_{\rm int} = \alpha_1 J_{1}^x + \alpha_2 
		J_{2}^x + \frac{\alpha_{12}}{j} J_{1}^xJ_{2}^x.
	\end{aligned}
\end{equation}
Here, $J_m=(J_{m}^x, J_{m}^y, J_{m}^z)$ for $m=1,2$ denote the collective 
spin operators. The primary resonance condition is imposed by setting 
$\beta_m=4\pi j$, and the system is initialized in the state 
$|\psi(0)\rangle=|J_{1}^z=-j\rangle \otimes |J_{2}^z=-j\rangle$. The 
time-evolved state is thus obtained as $|\psi(\tau)\rangle=U^{\tau} 
|\psi(0)\rangle$. 

\begin{figure}[htbt]
	\centering
	\includegraphics{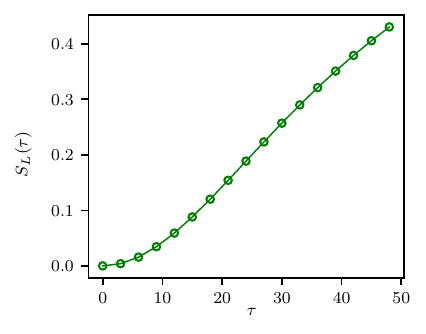}
	\caption{The growth of linear entropy is shown for the coupled kicked 
	top. The parameters are considered as $j=50, \alpha_1= 1,\alpha_2= 0.05, 
	\alpha_{12}= 0.03$.}
	\label{fig:S_top}
\end{figure}

As in the interacting kicked rotor, the coupled kicked top exhibits finite 
entanglement growth. We quantify this entanglement using the linear entropy,
\begin{equation}
	S_L(\tau) = 1 - \Tr[\rho_1(\tau)^2],
\end{equation}
where $\rho_1(\tau) = \Tr_2[\ket{\psi(\tau)}\bra{\psi(\tau)}]$ is the reduced 
density matrix of the first subsystem. The time evolution of this linear 
entropy is plotted in Fig.~\ref{fig:S_top}.


\end{document}